\documentclass[cits]{PoS}

\usepackage{graphicx}% 
\usepackage{epsf}
\usepackage{epsfig}

\usepackage{amsmath}
\usepackage{amsfonts,amssymb}
\usepackage{bm}% bold math

\usepackage{dsfont}
\usepackage{slashed}
\usepackage{booktabs}
\usepackage{placeins}

\newcommand{\cl}{\centerline}

\newcommand{\be}{\begin{equation}}
\newcommand{\ee}{\end{equation}}
\newcommand{\bea}{\begin{eqnarray}}
\newcommand{\eea}{\end{eqnarray}}

\newcommand{\Dlr}{\buildrel \leftrightarrow \over D\raise-1pt\hbox{}}

\title{Hadron Structure}

\ShortTitle{Hadron Structure}

\author{\speaker{Martha Constantinou}\\
         Department of Physics, University of Cyprus, P.O. Box 20537, 1678 Nicosia, Cyprus\\
       E-mail: \email{marthac@ucy.ac.cy}}

\abstract{This is a review of recent developments in hadron
structure within the framework of Lattice QCD. The main focus is
on recent achievements in the evaluation  of nucleon quantities, such
as the axial charge, electromagnetic form factors, the Dirac and Pauli
radii, the quark momentum fraction and the spin content of the
nucleon, in view of simulations at pion masses very close to their
physical value. A discussion of the systematic uncertainties and the
computation of the disconnected contributions using dynamical
simulations is also included. Results emerging the properties of
particles other than the nucleon are summarized, highlighting selected
hyperon and meson form factors. }

\FullConference{32st International Symposium on Lattice Field Theory - LATTICE 2014\\
		June 23 - June 28, 2014\\
		New York, USA}

\begin{document}

\section{Introduction}

Recent progress in the numerical simulation of Lattice QCD has been
impressive. This has been  due to  improvements in the algorithms and the development of
new techniques, as well as,  the increase in computational power, that
have enabled simulations to be carried out at parameters very close to
their physical values. The role of Lattice QCD is twofold: to make
contact with well determined experimental quantities with their
ab initio calculation, as well as, to make predictions on quatities 
that are not easily accessible in experiment, providing input to
phenomenology as well as new input for beyond the Standard Model
Physics.

Understanding nucleon structure from first principles is considered a
milestone of hadronic physics and numerous experiments have been
devoted to its study, starting with the measurements of the
electromagnetic form factors initiated more than 50 years
ago. Reproducing these key observables within the Lattice QCD 
formulation is a prerequisite to obtaining reliable predictions on
observables that explore Physics beyond the Standard Model. There is a
rich experimental program in major facilities (JLab, MAMI, MESA, etc)
investigating hadron structure, such are the origin of the nucleon spin, the
proton radius puzzle, and searching for new Physics, like $(g-2)_\mu$
and dark photon searches.

The 12~GeV upgrade of the Continuous Electron Beam
Accelerator Facility  at JLab~\cite{Jlab12}  will allow to employ new methods for
studying the basic properties of hadrons. Hadron structure has
been an essential part of the Physics Program which involves new and
interesting high precision experiments, such as nucleon resonance
studies with CLAS12, the longitudinal spin structure of the nucleon,
meson spectroscopy with low $Q^2$ electron scattering, the nucleon
generalized parton disctribution functions, high precision measurement
of the proton charge radius, and many more.

The experiments on the proton radious have attracted a lot of interest
since accurate measurements of the root mean square (r.m.s) charge radius from muonic
hydrogen~\cite{Antognini} ($\langle r^2_p \rangle_{\mu H} = 0.84\,{\rm fm}^2$) 
is 7.7$\sigma$ yield a value smaller that the radius determined from elastic e-p scattering and
hydrogen spectroscopy ($\langle r^2_p \rangle_{e p} =0.88\,{\rm fm}^2$)
\cite{Mohr:2012tt}. The 4$\%$ difference in the two measurements
is currently not explained. We note that the measurements in the muonic hydrogen
experiments are ten times more accurate than other measurements and
they are very sensitive to the proton size. In particular, the radius
is measured from the energy difference between the 2P and 2S states of
the muonic hydrogen~\cite{Pohl2010}
and more accurate esxperiments are planned at PSI.

Another interesting topic is the elastic light-by-light scattering
($\gamma\,\gamma \to \gamma\,\gamma$ ) which was never observed
directly, but only indirectly by its effects on the anomalous magnetic
moments of electrons and muons~\cite{lightlight}. In addition, 
photon-photon collisions in 'ultraperipheral' collisions of proton and
lead beams have been detected. Recently, there has been a study confirming that
light-by-light scattering could be directly detected at LHC, for the
first time, at 5.5-14 TeV~\cite{d'Enterria:2013yra}, due to the large
'quasireal' photons fluxes in electromagnetic interactions of protons
and lead ions.

The above few examples illustrate that hadron structure is a very rich
field of research relevant  to new Physics searches. Thus, Lattice QCD
does not only provide input to on-going experiments, but also gives
guidance to new searches with high-credibility results.

In these proceedings we discuss  representative observables probing
hadron structure for which there has been recent activity. Topics to
be covered include benchmark quantities, such as the nucleon axial charge,
electromagnetic form factors, the Dirac and Pauli radii, the quark
momentum fraction, as well as the nucleon spin, including disconnected
contributions. The systematic uncertainties are investigated and where
possible we compare with experimental / phenomenological data. 
 Recent results on Generalized Form Factors for
other baryons and mesons are also presented, as well as, 
perspectives and future directions.

\section{Nucleon Sector}
\vskip -0.2cm

Although the nucleon is the only stable hadron in the Standard Model,
its structure is not fully understood. Being one of the
building-blocks in the universe, the nucleon provides an extremely
valuable laboratory for studying strong dynamics providing important
input that can also shed light in new Physics searches. There have
been numerous recent lattice QCD results   on nucleon
observables. Here, we discuss selected  achievements, as well as,
challenges involved in these computations. 

\begin{figure}[!h]
\cl{\includegraphics[scale=0.5]{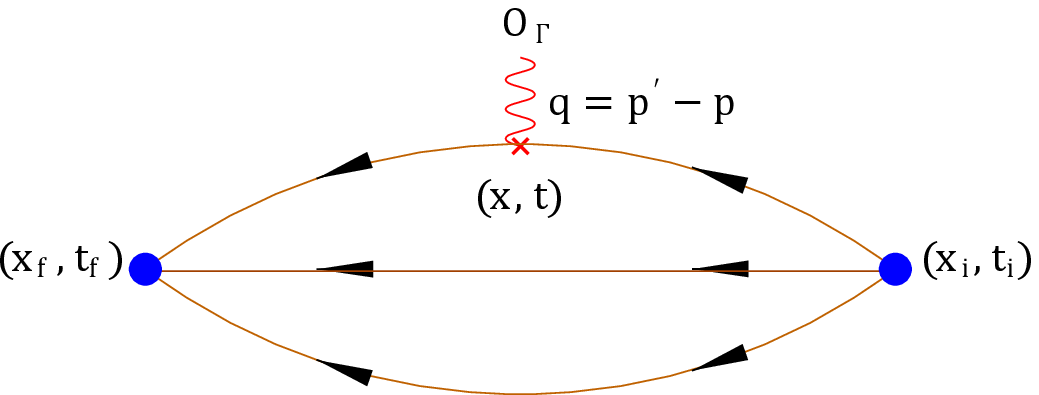} \hspace{1cm}
\includegraphics[scale=0.5]{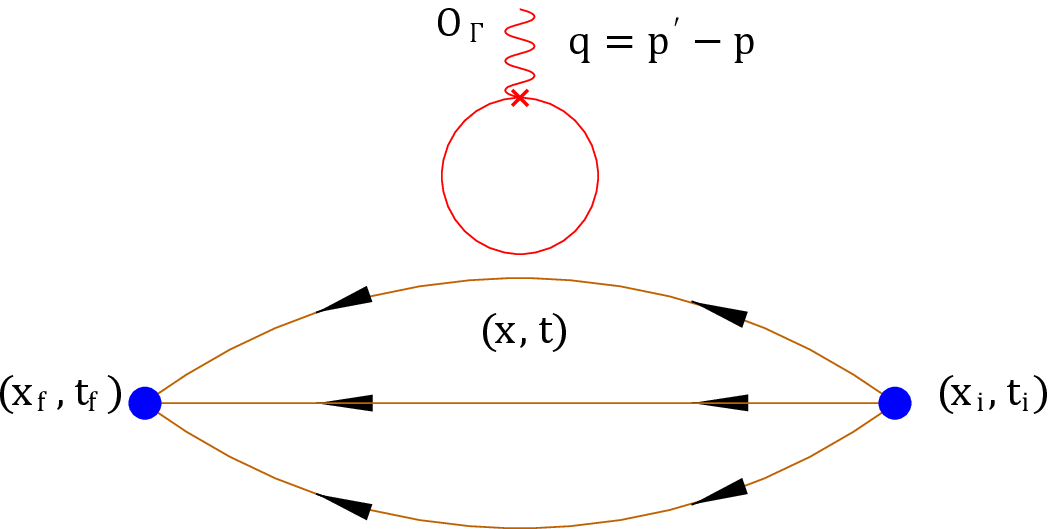}}
\vskip -0.1cm
\caption{Connected (left) and disconnected (right) contributions to
the nucleon three-point function.}
\label{fig1}
\end{figure}

 In a nutshell, in the evaluation of nucleon matrix elements in lattice QCD
 there are two type of diagrams
entering 
shown in Fig.~\ref{fig1}. The disconnected diagram has been neglected
in most of the studies because it is very noisy and expensive to
compute. During the last few years a number of groups are studying various
techniques for its computation and first results already appear in the
literature
\cite{Babich:2010at, QCDSF:2011aa, Engelhardt:2012gd, Abdel-Rehim:2013wlz}.
\vskip -0.52cm
\FloatBarrier
In the computation of nucleon matrix elements one needs  
appropriate two- and  three-point correlation functions defined as: 
\vspace{-0.3cm}
\bea
 G^{2pt}(\vec q, t_f) &=& \sum_{\vec x_f} \, e^{-i\vec x_f \cdot \vec q}\,
\Gamma^0_{\beta\alpha}\, \langle {{J_{\alpha}(\vec x_f,t_f)}}{{\overline{J}_{\beta}(0)}} \rangle \,, \\
 G^{3pt}_{\cal O}(\Gamma^\mu,\vec q, t_f) &=& \sum_{\vec x_f, \vec x} \, e^{i\vec x \cdot \vec q}\,e^{-i\vec x_f \cdot \vec p'}
     \Gamma^\mu_{\beta\alpha}\, \langle { J_{\alpha}(\vec x_f,t_f)} {\cal O}(\vec x,t) {\overline{J}_{\beta}(0)} \rangle\,.
\eea
\vskip -0.2cm
\noindent
The  projectors $\Gamma^\mu$ are defined as
$\Gamma^0 \equiv \frac{1}{4}(1+\gamma_0),\,\, \Gamma^k \equiv \Gamma^0\cdot\gamma_5\cdot\gamma_k\,$.
Other $\Gamma$-variations can be employed, in order to compute the quantities of interest. The
lattice data are extracted from dimensionless ratio of the two- and
three-point correlation functions:
\vspace{-0.11cm}
\bea
R_{\cal O}(\Gamma,\vec q, t, t_f) \hspace*{-.2cm}&{=}&\hspace*{-.2cm} \frac{G^{3pt}_{\cal O}(\Gamma,\vec q,t)}{G^{2pt}(\vec 0, t_f)}
\hspace*{-0.1cm}\times\hspace*{-0.1cm}\sqrt{\frac{G^{2pt}({-}\vec q, t_f{-}t)G^{2pt}(\vec 0, t)G^{2pt}(\vec 0, t_f)}{G^{2pt}(\vec 0  , t_f{-}t)G^{2pt}({-}\vec q,t)G^{2pt}({-}\vec q,t_f)}}\,\,
{\rightarrow \atop {{t_f{-}t\rightarrow \infty} \atop {t{-}t_i\rightarrow \infty}}}\,\,
\Pi (\Gamma,\vec q) \,.
\label{EqRatio}
\eea
\vskip -0.11cm
\noindent
The above ratio is considered optimized since it does not contain
potentially noisy two-point functions at large separations and because
correlations between its different factors reduce the statistical
noise. The most common method to extract the desired matrix element is to 
look for a
plateau with respect to the current insertion time, $t$ (or,
alternatively, the sink time, $t_f$), which should be located at a
time well separated from the creation and annihilation times in order
to ensure single state dominance. To establish proper connection to
experiments we apply renormalization which, for the quantities
discussed in this review, is multiplicative:
\vspace{-0.2cm}
\be
\Pi^R (\Gamma,\vec q) = Z_{\cal O}\,\Pi (\Gamma,\vec q)\,.
\ee
\vskip -0.2cm
Finally, the nucleon matrix elements can be parameterized in terms of
Generalized Form Factors (GFFs). As an example we take the axial current
insertion which decomposes into two Lorentz invariant Form Factors
(FFs), the axial ($G_A$) and pseudoscalar ($G_p$):
\vspace{-0.3cm}
\be
\hspace{-0.4cm}
\langle N(p',s')|\bar\psi(x)\,\gamma_\mu\,\gamma_5\,\psi(x)|N(p,s)\rangle= i \Bigg(\frac{
            m_N^2}{E_N({\bf p}')E_N({\bf p})}\Bigg)^{1/2} \hspace{-0.25cm}
            \bar{u}_N(p',s') \Bigg[
            G_A(q^2)\gamma_\mu\gamma_5
            +\frac{q_\mu \gamma_5}{2m_N}G_p(q^2)
            \Bigg]u_N(p,s)\,,
\label{axial_decomp}
\ee
\vskip -0.1cm
\noindent
where $q^2$ is the momentum transfer in Minkowski space (hereafter,
$Q^2=-q^2$).

In these proceedings I will mostly consider the flavor isovector
combination for which the disconnected contribution cancels
out; strictly speaking, this happens for actions with  exact isospin
symmetry. Another advantage of the isovector combination is that the
renormalization simplifies considerably.

\subsection{Nucleon Axial Charge}
One of the fundamental nucleon observables is the axial charge, 
$g_A \equiv G_A(0)$, which is determined from the  forward matrix
element of the axial current. $g_A^q$ gives the intrinsic quark  spin
in the nucleon. It governs the rate of $\beta$-decay and has been
measured precisely. In the lattice QCD it can be determined directly
from the evaluation of the matrix element and thus, there is no
ambiguity asocciated to fits. For this reasons, $g_A$ is an optimal
benchmark quantity for hadron structure computations. It is thus
essential for  lattice QCD to reproduce its experimental value or
if a deviation is observed to understand its origin. 
\begin{figure}[!h]
\cl{\includegraphics[scale=0.39]{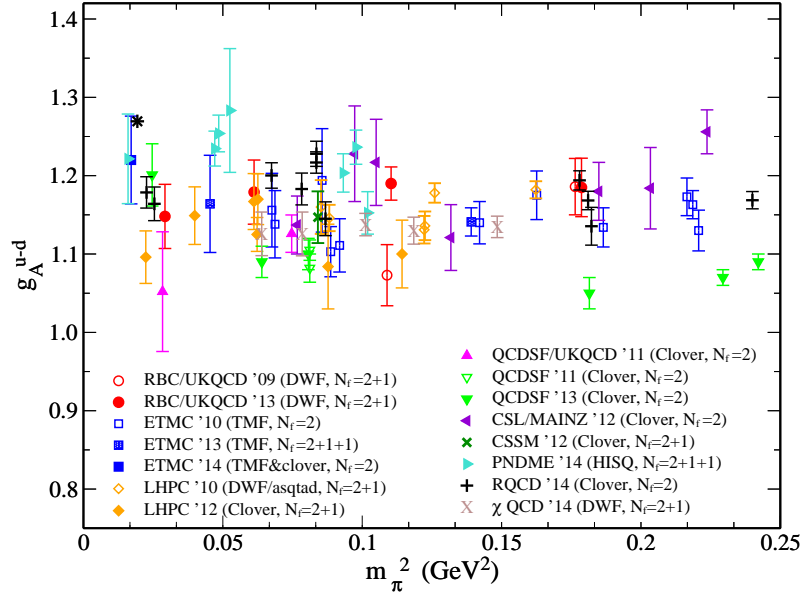}}
\caption{Collection of lattice results for $g_A$. In chronological
order these correspond to: $N_f{=}2{+}1$ DWF
(RBC/UKQCD~\cite{Yamazaki:2008py, Yamazaki:2009zq},
RBC/UKQCD~\cite{Ohta:2013qda}, $\chi$QCD~\cite{chiQCD14}),
$N_f{=}2{+}1$ DWF on asqtad sea (LHPC~\cite{Bratt:2010jn}), $N_f{=}2$
TMF (ETMC~\cite{Alexandrou:2010hf}), $N_f{=}2$ Clover
(QCDSF/UKQCD~\cite{Pleiter:2011gw}, CLS/MAINZ~\cite{Capitani:2012gj},
QCDSF~\cite{Horsley:2013ayv}, RQCD~\cite{Bali:2013nla,RQCD14}),
$N_f{=}1{+}2$ Clover (LHPC~\cite{Green:2012rr},
CSSM~\cite{Owen:2012ts}), $N_f{=}2{+}1{+}1$ TMF
(ETMC~\cite{Alexandrou:2013joa}), $N_f{=}2{+}1{+}1$ HISQ
(PNDME~\cite{Bhattacharya:2013ehc,PNDME14}), $N_f{=}2$ TMF with Clover
(ETMC~\cite{ETMC14}). The asterisk is the experimental value.}
\label{fig2}
\end{figure}
\FloatBarrier
There are numerous computations from many collaborations
and some selected results are shown in Fig.~\ref{fig2} as a function
of the squared pion mass. These results have been obtained using dynamical 
gauge field configurations with ${\cal O}(a)$-improved lattice QCD
actions,  namely Domain Wall Fermions (DWF), Staggered, Clover,
Twisted Mass Fermions (TMF) and HISQ fermions~\cite{Yamazaki:2008py,
  Yamazaki:2009zq, Bratt:2010jn, Alexandrou:2010hf, Pleiter:2011gw,
  QCDSF:2011aa, Capitani:2012gj, Green:2012rr, Owen:2012ts,
  Horsley:2013ayv, Alexandrou:2013joa, Bhattacharya:2013ehc,
  Ohta:2013qda, Bali:2013nla}. For a fair comparison we include only
results obtained from the plateau method without any volume
corrections. The latest achievement of the Lattice Community are the
results at  the physical point for which there is no necessity of
chiral extrapolation eliminating an up to now uncontrolled
extrapolation. The ones at the two lowest values of the pion
mass correspond to PNDME (128~MeV)~\cite{PNDME14} and ETMC
(130~MeV)~\cite{ETMC14}, and are in agreement with the
experimental value: $g_A^{\rm exp}=0.2701(25)$~\cite{PDG12}. 
Of course the statistical errors are still large and it is necessary
to increase the statistics and study the volume and lattice spacing
dependence before finalizing these results. One should also keep in
mind that the results shown in  Fig.~\ref{fig2} are at a given lattice
spacing and volume, and each Collaboration addresses systematic effects
in different ways. We will comment later on this issue.

\begin{figure}[h!]
\cl{\includegraphics[scale=0.25]{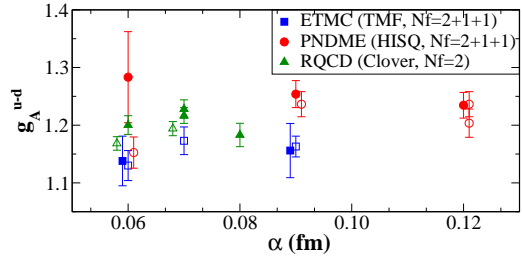}}
\vskip -0.1cm
\caption{The axial charge as a function of the lattice spacing. The data
correspond to: a. TM fermions (ETMC): open/filled symbols for $m_\pi\sim465\,/260$ MeV,
b. HISQ fermions (PNDME): open/filled symbols for $m_\pi\sim310\,/220$ MeV,
c. Clover fermions (RQCD): open/filled symbols for
$m_\pi\sim420\,/280$ MeV. Points have been horizontally shifted for clarity.}
\label{fig3}
\end{figure}

\vskip 0.5cm
$\bullet$ {\it{Cut-off effects}}:
For a proper continuum extrapolation one requires three lattice
spacings, which is computationally very costly, especially as we
approach the physical point. Thus, it is crucial to investigate cut-off
effects using an ensemble with a heavier pion mass. Such a study was
done with TMF~\cite{Alexandrou:2010hf}, Clover~\cite{RQCD14} and
HISQ~\cite{PNDME14} fermions and the results are shown in
Fig.~\ref{fig3}. Data with the same symbol correspond to similar pion
mass, spanning from $\sim$220 to 480 MeV. These results show that with
lattice spacings up to about 0.1~fm lattice artifacts are small
compared to the statistical accuracy. 
\FloatBarrier
\vskip 0.5cm
$\bullet$ {\it{Excited states contamination}}:
The interpolating field used to create a hadron of given quantum numbers
 couple in addition to the excited states.
While for two-point functions identification of the
ground is straight forward for three-point functions it is more saddle.
The most common approach is the so called plateau
method in which one probes the large Euclidean time evolution of the
ratio in Eq.~(\ref{EqRatio})
\vspace{-.2cm}
\be
R_{\cal O}(\Gamma,t_i,t,t_f) {\rightarrow \atop {{(t_f{-}t)\,\Delta >>1} \atop {(t{-}t_i)\,\Delta >>1 }}}\,\,
{\cal M}\Bigg[ 1 {+} \alpha\,e^{-(t_f{-}t)\,\Delta(p')} {+} 
\beta\,e^{-(t{-}t_i)\,\Delta(p)} {+}\cdots \Bigg]\,.
\ee
As can be seen in the above equation, the excited states contributions
fall exponentially with the sink-insertion ($t_f-t$) and
insertion-source ($t-t_i$) time separation. So, it is possible to
reduce the unwanted excited states contamination by increasing the source-sink
separation, but this comes with a cost of increased statistical noise.  

Alternatively, the matrix element may be obtained by performing a
2-state fit to account for contributions from the first excited state. One can also 
perform simultaneous fits on data for different sink-source separations 
\vspace{-.1cm}
\begin{align}
{\cal C}^{(3),T}_{\Gamma}&(t_i,t,t_f;\vec{p}_i,\vec{p}_f) \approx 
        |{\cal A}_0|^2 \langle 0 | O_\Gamma | 0 \rangle  e^{-M_0\,(t_f-t_i)} +
        |{\cal A}_1|^2 \langle 1 | O_\Gamma | 1 \rangle  e^{-M_1\,(t_f-t_i)} +{}\nonumber\\[2ex]
      & {\cal A}_0{\cal A}_1^* \langle 0 | O_\Gamma | 1 \rangle  e^{-M_0 (t-t_i)} e^{-M_1 (t_f-t)} +
       {\cal A}_0^*{\cal A}_1 \langle 1 | O_\Gamma | 0 \rangle  e^{-M_1 (t-t_i)} e^{-M_0 (t_f-t)} 
\end{align}
\vskip -0.1cm
\noindent
where $M_0$ and $M_1$ are the masses of the ground and first excited
state, respectively, and ${\cal A}_0\,{\cal A}_1$ are the
corresponding amplitudes.

\begin{figure}[!h]
\cl{\includegraphics[scale=0.25]{./gA_vs_Tsink_RQCD_300MeV.eps}\quad 
    \includegraphics[scale=0.2]{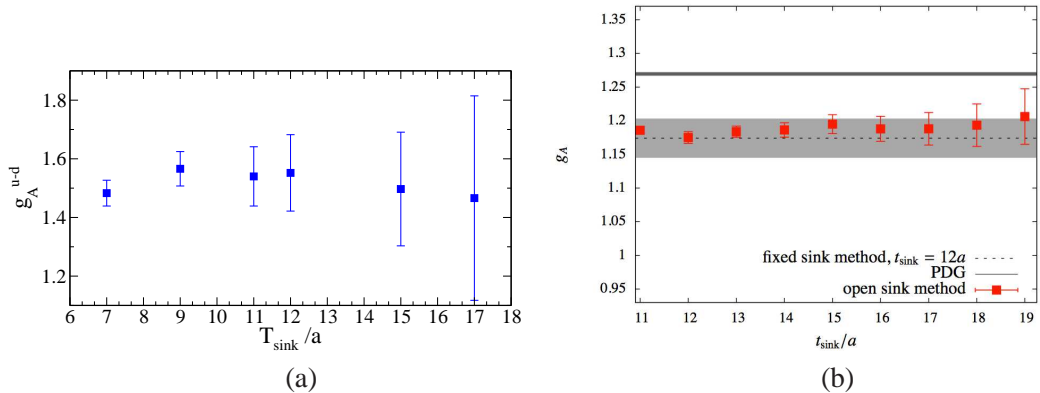}}
\vskip -0.05cm
\cl{\hspace{2cm} (a) \hspace{6.5cm} (b)\hspace{1cm}}
\vskip -0.15cm
\caption{The axial charge as a function of the
  source-sink separation for (a). $N_f{=}2$ Clover fermions at $m_\pi~300$
  MeV (RQCD~\cite{RQCD14}), and (b).  $N_f{=}2{+}1{+}1$ TMF (ETMC~\cite{Dinter:2011sg}).} 
\label{fig4}
\end{figure}

\begin{figure}[!h]
\cl{\includegraphics[scale=0.34]{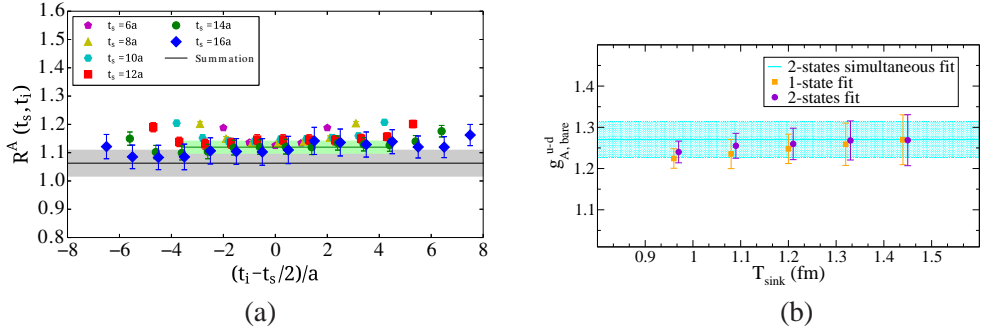}\quad
    \includegraphics[scale=0.21]{./gA_2state_fit_mpi310MeV_1306.5435_2.eps}}
\vskip -0.05cm
\cl{\hspace{2cm} (a) \hspace{6.5cm} (b)\hspace{1cm}}
\vskip -0.2cm
\caption{(a). The ratio of the axial charge for $N_f{=}2{+}1{+}1$ TMF
(ETMC~\cite{Alexandrou:2013joa}), (b).  Bare values for $g_A$ as
a function of the source-sink separation using $N_f{=}2{+}1{+}1$ HISQ
fermions at $m_\pi=310$ MeV.}
\label{fig5}
\end{figure}

A third method is the so-called  summation method in which we sum the
ratio from the source to the sink~\footnote{The result also holds by
excluding the source and sink points to avoid contact terms.}. This
way, the excited state contaminations are suppressed by exponentials
decaying with $(t_f- t_i)$ rather than $(t_f - t)$ and $(t - t_i)$. 
However, one needs the slope of the summmed ratio:
\vspace{-.1cm}
\be
\sum_{t=t_i}^{t_f}\,R(t_i,t,t_f) = {\rm const.} +
{\cal M}\, (t_f-t_i) + 
{\cal O}\Bigg( e^{-((t_f-t_i)\,\Delta(p'))}\Bigg) + 
{\cal O}\Bigg( e^{-((t_f-t_i)\,\Delta(p))} \Bigg) \,.
\ee
\vskip -.1cm
All the aforementioned methods have been applied in the extraction of
$g_A$ and some of the works are presented here. In a study by RQCD~\cite{RQCD14}
at $m_\pi~300$ MeV, no change of $g_A$ has been seen by varying the
separation from 0.5-1.2 fm (see Fig.~\ref{fig4}a); this is also
confirmed by a high precision study at $m_\pi\sim370$ MeV using TMF
\cite{Dinter:2011sg} in which the source-sink separation takes values
within the range 0.9-1.6 fm~\cite{Dinter:2011sg}. 
\FloatBarrier
Studying the summed ratio using TMF  no curvature
in the slope is observed and the result extracted for $g_A$  agrees with the plateau method as can be
seen in Fig~\ref{fig5}a. The 2-state analysis of
PNDME~\cite{Bhattacharya:2013ehc} shown in Fig.~\ref{fig5}b (purple
points) agrees with the corresponding single state fit (orange
points), as well as with the 2-state simultaneous fit using the lattice data on
different separations (blue band).
\vskip -0.2cm
\FloatBarrier
Recently the Feynman-Hellmann approach was utilized for the
computation of $g_A$ by CSSM/ QCDSF/ UKQCD~\cite{Chambers:2014qaa} 
at $m_\pi=470$ MeV, which is based on the introduction of an external
spin operator to the fermion part of the action:
$$S {\to} S(\lambda) {=} S {+} \lambda \sum_x \overline{q}(x) i\gamma_5\gamma_3 q(x).$$
The axial charge can then be extracted by the linear response of the
hadron energies:
\be
\Delta q {=} \left. \frac{\partial E(\lambda)} {\partial \lambda}
 \right|_{\lambda=0} {=} \frac{1}{2 M} \langle N | \overline{q}
  i\gamma_5\gamma_3 q | N \rangle\,.
\ee
The authors find improved statistical accuracy compared to the
standard method. It would be
interesting to test this approach in observables highly contaminated
by excited states, since it is believed that this method is less
susceptible to excited states contributions.

\bigskip
We note that  all high statistics studies of systematic uncertainties have been
performed at relatively large values of the pion mass. It
is thus essential  to also perform
similar investigations at values of the pion mass closer to the physical one.
Given that the signal to noise error decreases exponentially as
the pion mass decreases:
%\be
%\sim \sqrt(N_{\rm meas})\times e^{-\left(m_N + 3\,m_\pi/2 \right)}
%\ee
%($N_{\rm meas}$: number of independent measurements, $m_N$: nucleon
%mass, $m_\pi$: pion mass)  
  one needs to
increase considerably the number of independent measurements leading
to increase computational cost. 
Thus,
noice reduction methods are highly valuable. RBC/UKQCD has implemented
a new class of covariant approximation averaging techniques, called
 All-Mode-Averaging (AMA)~\cite{Shintani:2014vja} which allows for a
significant  decrease of the statistical error over conventional
methods at reduced computational cost. 
 The AMA
technique achieves an increase of the number of measurement,  $N_{\rm meas}$, at a low computational
cost, by constructing an improved operator which is built from 
a low-precision and thus cheap  part, and a correction term, to compensate the
bias introduced by the approximate measurements:
\be
\langle {\cal O}^{\rm impr} \rangle = \langle{\cal O}^{\rm
  approx}\rangle + \langle{\cal O}^{\rm rest}\rangle\,,\quad 
{\cal O}^{\rm rest}={\cal O}^{\rm exact}-{\cal O}^{\rm approx}\,.
\ee
The approximate operator has the same covariance properties as
the original one, but a much smaller construction cost and thus 
${\cal O}^{\rm impr}$ has smaller statistical errors without
additional computational cost.
In other words, the AMA allows to increase statistics using  a large number of sloppy measurements and a small number of exact ones:
\be
O_{\rm AMA} = \frac{1}{N_{\rm apprx} }\sum_{i=1}^{N_{\rm apprx}} O_{\rm apprx}^i + 
\frac{1}{N_{\rm exact}} \sum_{j=1}^{N_{\rm exact}} \left ( O_{\rm exact}^j - O_{\rm apprx}^j \right )\,.
\ee
It was shown that the error of the combined result depends
highly on the number of approximate measurements. For the case of
$g_A$ a speed up of $\sim$ 5-20 times is reported using $N_f{=}2{+}1$
DWF configurations on a $24^3\times48$ lattice, and up to 10-30 times
for a larger lattice, $32^3\times64$~\cite{Shintani:2014vja}.

\begin{figure}[!h]
\cl{\includegraphics[scale=0.3]{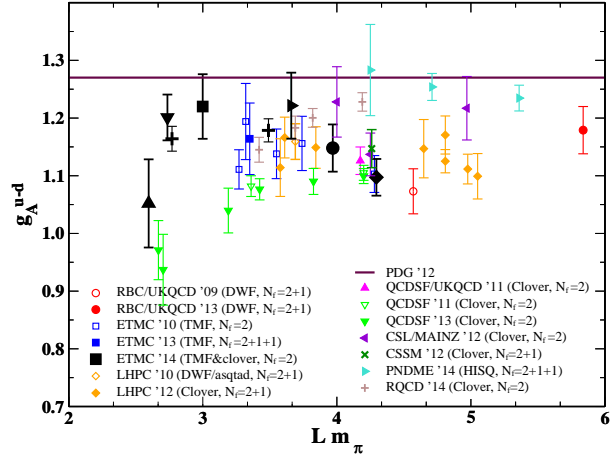}}
\caption{Summary of lattice results for $g_A$ as a function of the
finite scaling parameter $L\,m_\pi$. See caption of Fig. 2 for
references.} 
\label{fig6}
\end{figure}

\bigskip
$\bullet$ {\it{Finite volume effects}}:
To assess volume effects we plot $g_A$ against  $L\,m_\pi$,  in
Fig.~\ref{fig6}. In order for the data to be distinguishable we
restrict to pion mass of about 300~MeV. We highlight results at almost
physical values of the pion mass by the  black filled symbols. These
come from a range of collaborations, namely: PNDME~\cite{PNDME14} at
$m_\pi=128$~MeV,  ETMC~\cite{ETMC14} at $m_\pi=130$~MeV,
LHPC~\cite{Green:2012rr} at $m_\pi=149$ MeV,  RQCD~\cite{RQCD14} at
$m_\pi=150/157$~MeV, QCDSF~\cite{Horsley:2013ayv,Pleiter:2011gw} at
$m_\pi=170$~MeV and RBC~\cite{RBC14} at $m_\pi=170$ MeV. 
\FloatBarrier
One immediately observes that data at $L\,m_\pi\sim 3,\,3.7$ agree
with experiment, while data at $L\,m_\pi > 4$ are below the
experimental point. This behaviour is puzzling 
and further studies are required to clarify the volume dependence.  

\bigskip
In summary, based on current results on the axial charge,  
we conclude that cut-off effects are small, at least for $a
\le 0.1$ fm, and no indication of significant excited state
contamination has been observed indicating that sink-source time
separation of about 1~fm is sufficient. No clear conclusion can be
extracted regarding finite volume effects that need further
investigation. It is worth stressing however that the preliminary
value of $g_A$ determined by ETMC close to the physical point with
$L\,m_\pi \sim 3$ and $a<0.1$~fm is in agreement with the
experimental value.

\subsection{Nucleon Electromagnetic Form Factors}

\begin{figure}[!h]
\cl{\includegraphics[scale=0.6]{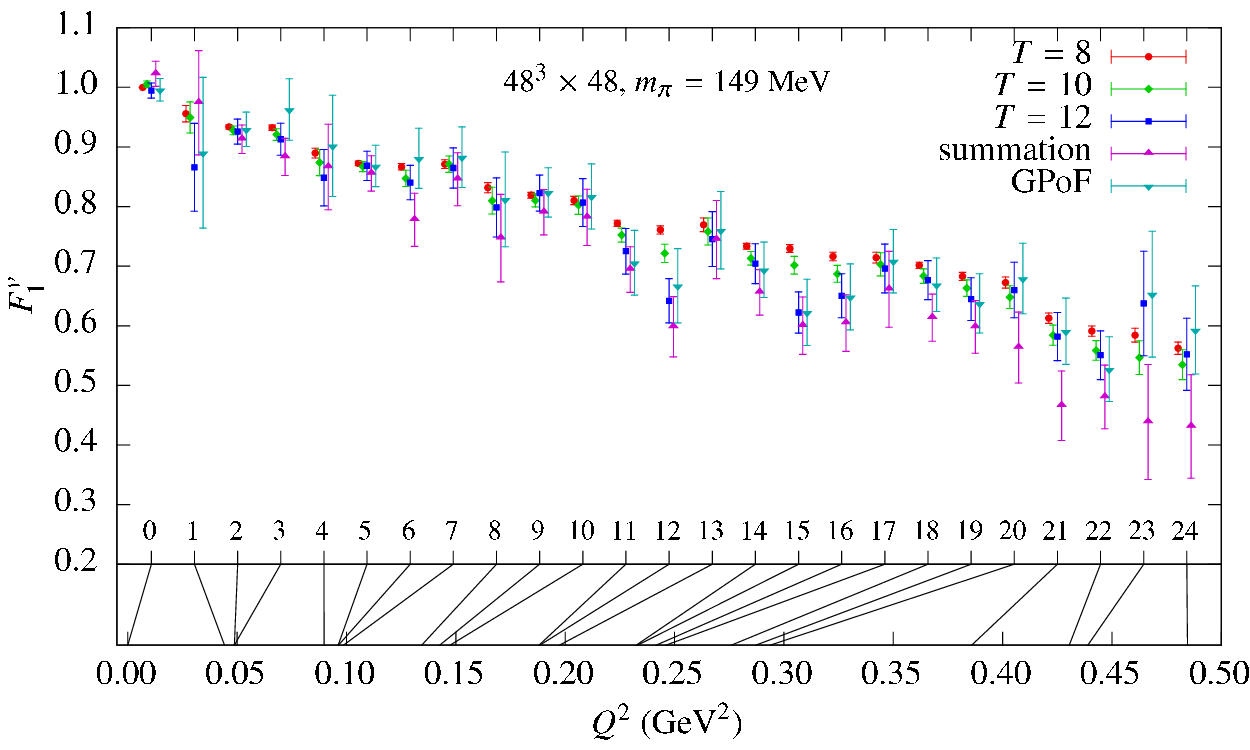}\,\,\,\,
    \includegraphics[scale=0.2]{./F1_F2_ratio.eps}}
\vskip -0.05cm
\cl{\hspace{2.5cm} (a) \hspace{7cm} (b)\hspace{1cm}}
\vskip -0.05cm
\caption{(a). Isovector $F_1$ extracted from different methods for
$N_f{=}2$ Clover fermions at $m_\pi=149$ MeV~\cite{Green:2014xba}
(b). Comparison of the Dirac and Pauli form factors for Clover 
and TM fermions at pion mass close to the physical point, plotted
against $Q^2$.The solid line is J. Kelly's parametrization of the
experimental data~\cite{Kelly}. }
\label{fig7}
\end{figure}

\begin{figure}[!h]
\cl{\includegraphics[scale=0.37]{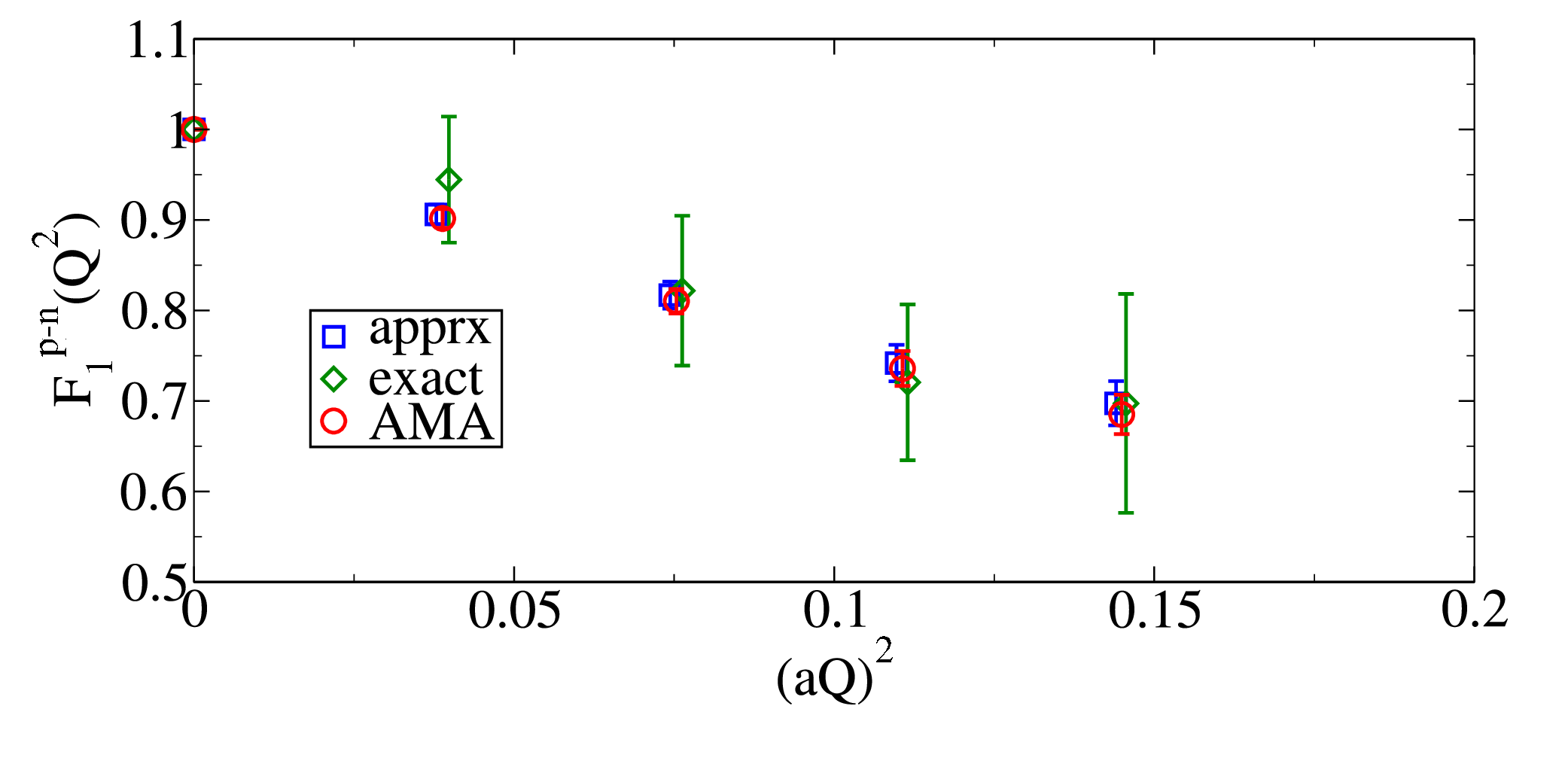}}
\vskip -0.3cm
\caption{The isovector Dirac form factor for $N_f{=}2{+}1$ DWF at
$m_\pi=170$ MeV~\cite{Lin:2014saa}. The points have been slightly
shifted for clarity.}
\label{fig8}
\end{figure}

The matrix element of the vector current decomposes into the Dirac and
Pauli FFs:
\be
\langle N(p',s')|\gamma_\mu|N(p,s)\rangle \sim \bar{u}_N(p',s') \Bigg[
 F_1(q^2)\gamma_\mu +F_2(q^2)\frac{i\,\sigma^{\mu\rho}\,q_\rho}{2m_N} \Bigg]u_N(p,s) \,.
\ee
The most recent study by LHPC at $m_\pi=149$ MeV, with statistics
exceeding 7000 measurements, explores three values of the source - sink
separation, on which the summation method is also applied. As an
example, we show the Dirac form factor $F_1(Q^2)$ in
Fig.~\ref{fig7}a, while in Fig.~\ref{fig7}b we compare results for
$F_1$ and $F_2$ using Clover fermions ($m_\pi=149$ MeV)~\cite{Green:2014xba} and
TMF ($m_\pi=130$ MeV)~\cite{ETMC14b}.
\FloatBarrier
\noindent 
We find a nice agreement between the two discretizations and the slope
has been improved compared to data at higher pion masses. Of course
the statistical errors are still large, and before reaching to
conclusions the statistics must be increased; the AMA technique could
be extremelly usefull in the error reduction, as can be seen in
Fig.~\ref{fig8}. The data correspond to $N_f{=}2{+}1$ DWF at
$m_\pi=170$ MeV~\cite{Lin:2014saa}. The utilization of the AMA
technique on a $32^3\times64$ lattice led to roughly a factor of 20
improvement in the computational efficiency, and to a dramatic
reduction in the statistical errors. 
\vskip -0.3cm
\FloatBarrier

Preliminary results for disconnected contributions are available by
LHPC~\cite{LHPC14} for the Sachs FFs:
\be
G_E(Q^2)=F_1(Q^2)-\frac{Q^2}{4m_N^2}F_2(Q^2), \quad G_M(Q^2)=F_1(Q^2)+F_2(Q^2)\,,
\ee
using hierarchical probing~\cite{Stathopoulos:2013aci} for
the computation of the quark loops. The gain of this method depends on
the observable and for the electromagnetic FFs is significant, leading
to results with very impresive accuracy at various values of the
momentum transfer. The hierarchical probing is a spatial dilution
method using a sequence of deterministic orthonormal vectors, called
Hadamard, built out of 1's and -1's in a specific order. The `level of
dilution' is increased gradually, at any stage of the computation,
while reusing existing data. The method improves the stochastic
estimator: ${\rm Tr}[A^{-1}] = E\{z^\dagger A^{-1} z\}$ ($z$: noise
vector), using the fact that $A^{-1}_{i,j}$ is dense but decays
exponentially as the distance, $k$, between the sites $i,\,j$, i.e. $|i-j|$
increases. The optimal distance $k$ for $A^{-1}_{i,j}\approx 0$ is
obtained using probing, where the results from level $i-1$ is used at
level $i$. The multi coloring of sites is done hierarchically and the
bias is removed by using a random starting vector. A factor of 10 speed up in the
computational cost over standard approach has been reported on a
$32^3\times 64$ Clover lattice at the strange quark mass.
 
\begin{figure}[!h]
\cl{\includegraphics[scale=0.57]{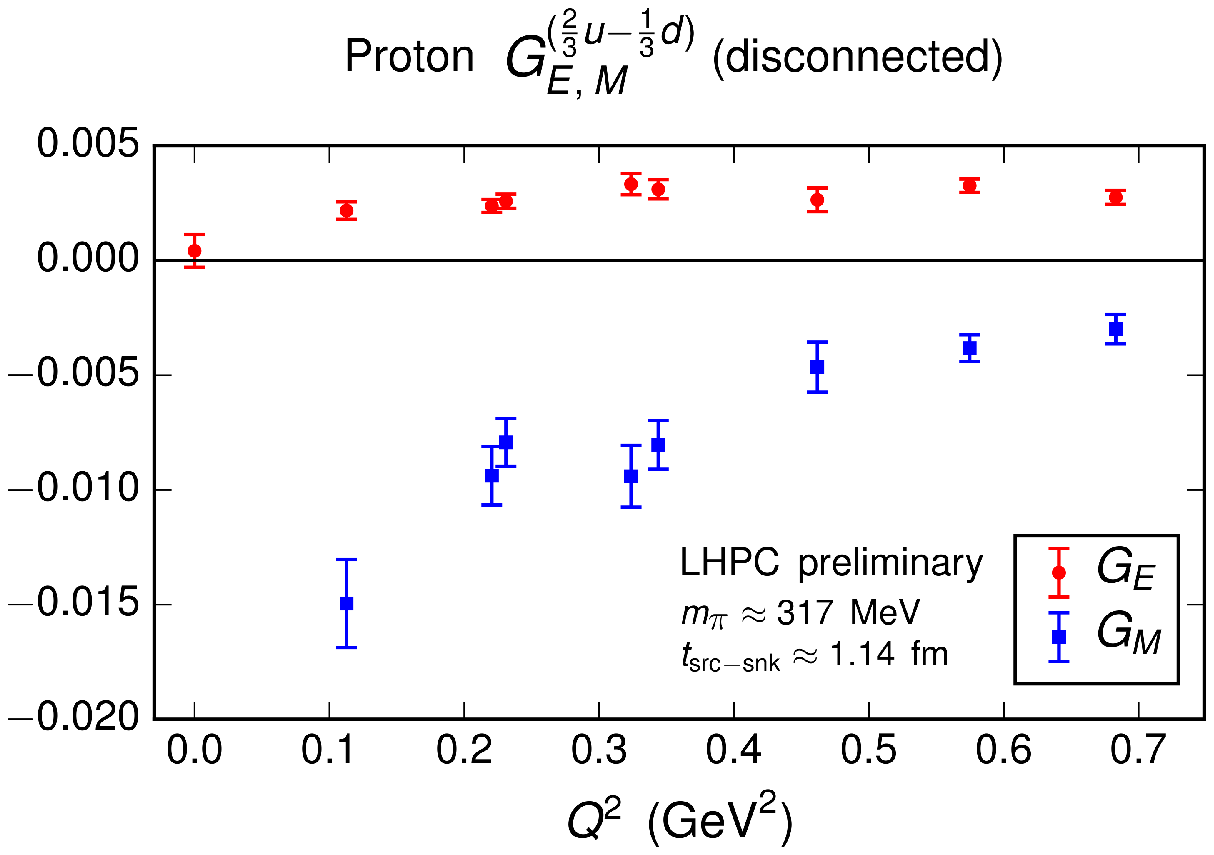}\,\,
    \includegraphics[scale=0.57]{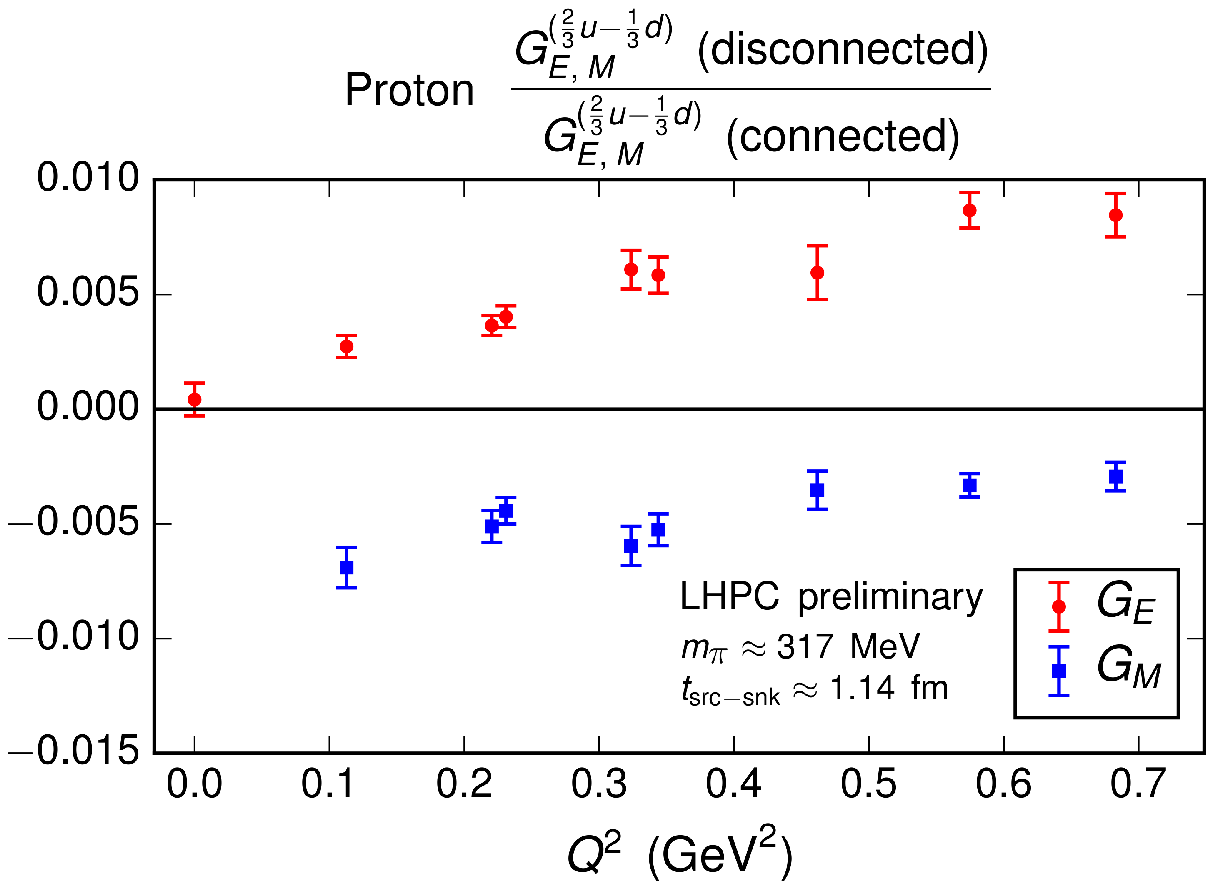}}
\vskip -0.4cm
\caption{Results for the proton Sach FFs using
$N_f{=}2{+}1$ Clover fermions at $m_\pi=317$ MeV~\cite{LHPC14}.}
\label{fig9}
\end{figure}

The connected and disconnected contributions to the proton Sachs FFs,
as well as their ratio are shown in Fig.~\ref{fig9} for $N_f{=}2{+}1$
Clover fermions at a pion mass of 317 MeV, and the statistics used is
$\sim{\cal O}(100\,000)$. It is interesting to see that the
disconnected contribution of $G_E$ increases the total value, while
the total $G_M$ decreases. In both cases, the disconnected
contribution are approximately 0.5$\%$ of the connected contribution.
\vskip -0.3cm
\FloatBarrier

\subsection{Dirac $\&$ Pauli radii}

\begin{figure}[!h]
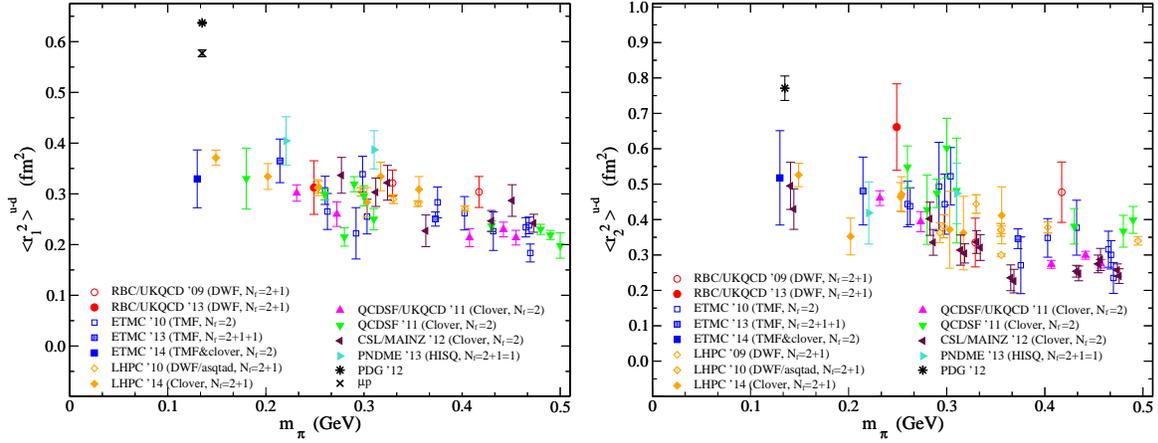

\cl{\includegraphics[scale=0.275]{./r1sq_vs_mpi.eps}\,\,
    \includegraphics[scale=0.275]{./r2sq_vs_mpi.eps}}
\vskip -0.2cm
\caption{Dirac (left) and Pauli (right) radii as a function of the
pion mass. The lattice data correspond to the following
discretizations: $N_f{=}2{+}1$ DWF (RBC/UKQCD~\cite{Yamazaki:2009zq,
Lin:2013bxa}), $N_f{=}2$ TMF (ETMC~\cite{Alexandrou:2011db}),
$N_f{=}2+1$ DWF on asqtad sea (LHPC~\cite{Bratt:2010jn}), $N_f{=}2$
Clover (QCDSF/UKQCD~\cite{Pleiter:2011gw},
QCDSF~\cite{Collins:2011mk}, CLS/MAINZ~\cite{Capitani:2012ef}),
$N_f{=}2{+}1{+}1$ TMF (ETMC~\cite{Alexandrou:2013joa}),
$N_f{=}2{+}1{+}1$ HISQ (PNDME~\cite{Bhattacharya:2013ehc}),
$N_f{=}2+1$ Clover (LHPC~\cite{Green:2014xba}), $N_f{=}2$ TMF with
Clover (ETMC~\cite{ETMC14b}). The experimental points have been taken
from Refs.~\cite{Antognini,Mohr:2012tt}.}
\label{fig10}
\end{figure}

In the non-relativistic limit the slope of the electromagnetic form
factors, $F_1,\,F_2$, at zero momentum transfer is characterized by the Dirac,
$r_1$, and Pauli, $r_2$, radii of charge:
\vspace{-0.10cm}
\be
F_i(Q^2) \sim F_i(0)\left(1-\frac{1}{6}\,Q^2 \langle r_i^2 \rangle + {\cal O}(Q^4) \right)\,, \qquad 
\langle r_i^2\rangle = -\frac{6}{F_i(Q^2)}\,\frac{dF_i(Q^2)}{dQ^2}\Big{|}_{Q^2=0}\,.
\ee
\vskip -0.12cm
\noindent
To determine the slope we fit the form factors using a dipole
function, and therefore, the r.m.s radii can be obtained from the
values of the dipole mass:
\vspace{-0.12cm}
\be
F_i(Q^2) \sim {F_i(0)}/{\left(1+{Q^2}/{m_i^2} \right)^2}\,, \qquad 
\langle r_i^2\rangle = {12}/{m_i^2}\,.
\ee
\vskip -0.12cm
\noindent
Input on these radii from Lattice QCD is significant due to
the persisting descrepancy between experiments involving electrons and
muons~\cite{Antognini,Mohr:2012tt}. A collection of lattice results
for $\langle r_1^2 \rangle$ appears in the left panel of Fig.~\ref{fig10} and for
$\langle r_2^2 \rangle$ in the right panel. Overall we find a good agreement among
different lattice discretizations and as we approach the physical
point we observe an increasing trend of the data which could be justified
be a logarithmic chiral behavior. The Pauli radius appears to have
larger statistical errors compared to the Dirac radius which is
expected since, in the dipole fit there is an additional parameter,
$F_2(0)$, that cannot be obtained directly from the lattice
data. Since the estimation of the radii is strongly dependent on the
small momenta, we need access to momenta closer to zero, and thus, 
simulations of nucleons of nucleons in larger boxes are necessary.

\FloatBarrier

\begin{figure}[!h]
\cl{\includegraphics[scale=0.41]{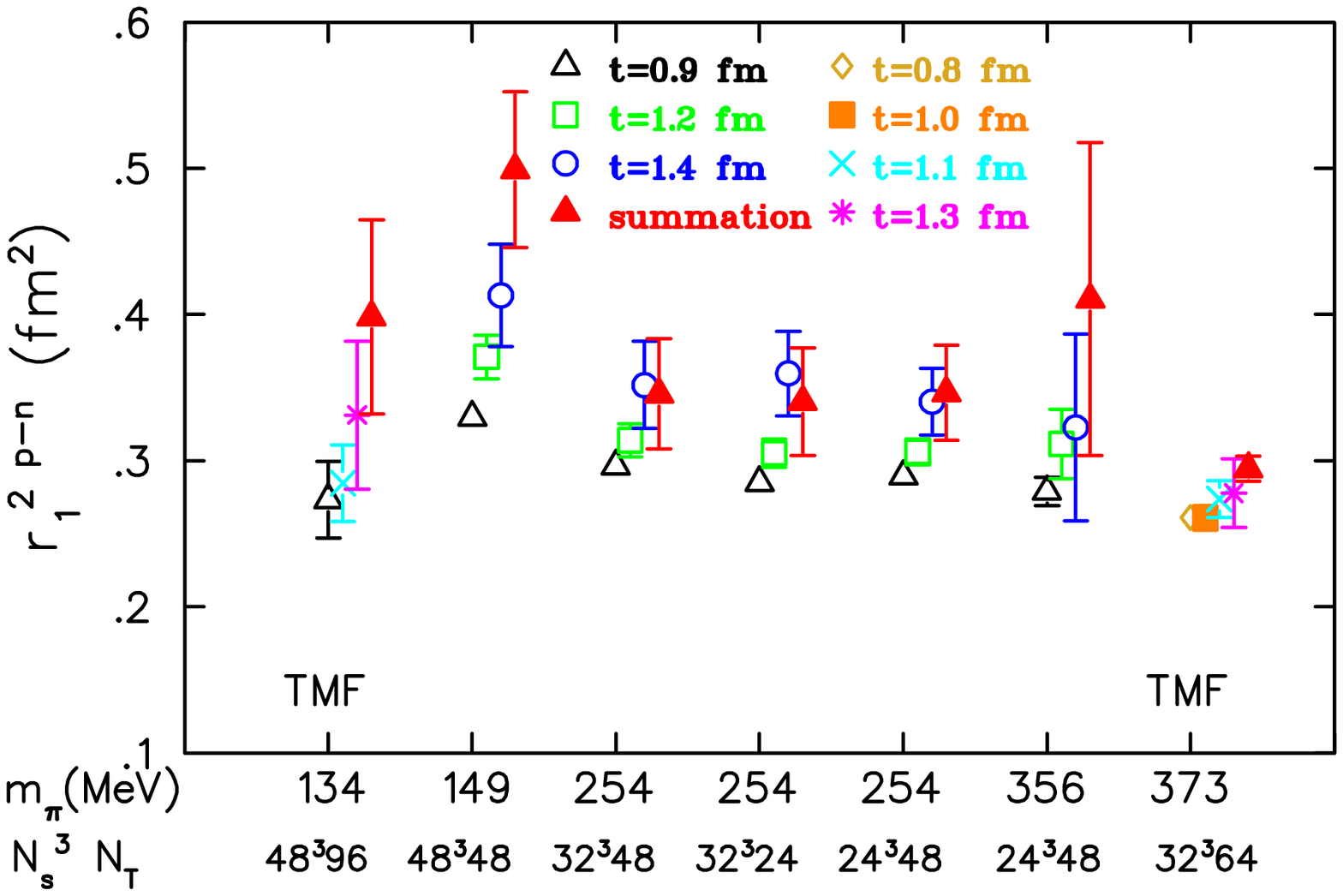}\quad
{\includegraphics[scale=0.6]{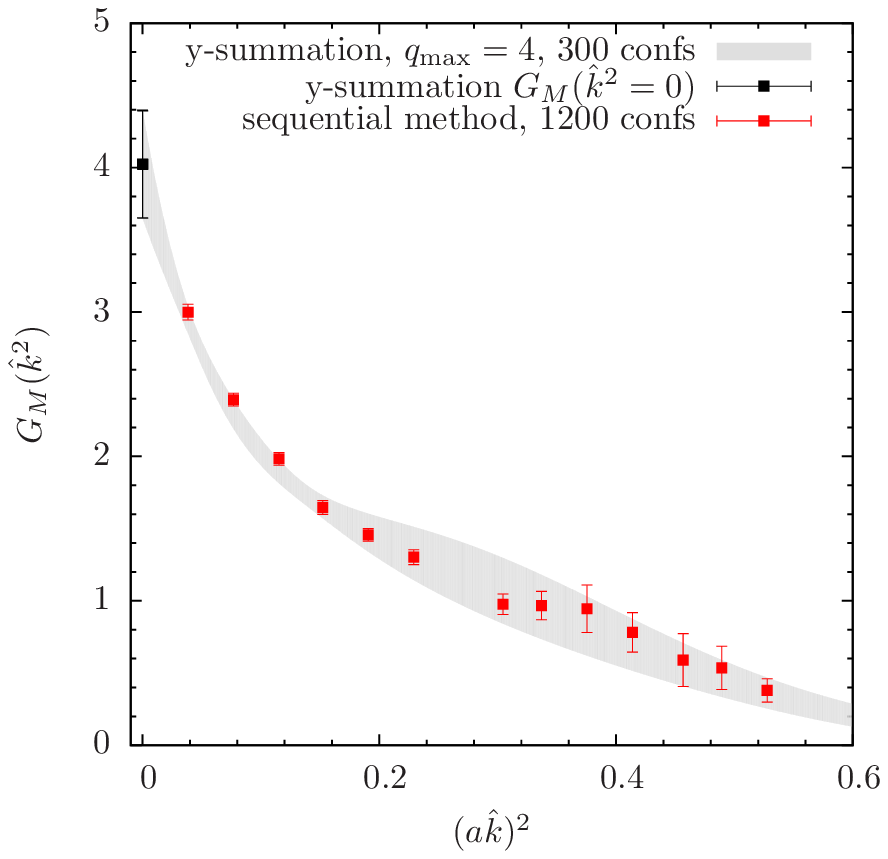}}}
\vskip -0.2cm
\caption{Left: Isovector $\langle r_1^2 \rangle$ for various ensembles and different
source-sink separations~\cite{Green:2014xba, ETMC14b}. Right:
Isovector $G_M$ for TMF extracted from a position space
method~\cite{ETMC14c}.}
\label{fig11}
\end{figure}

In the left panel of Fig.~\ref{fig11} we plot $r_1$ for a range of
pion mass as the sink-source separation increases. The study is
carried out  by LHPC \cite{Green:2014xba} and ETMC~\cite{ETMC14b} and
it include the result of the summation method. There is a clear
upward tendency as the sink-source time separation increases. Although
the results from the summation method agree with the value extracted
from the plateau method for the largest sink-source time separation,
the errors on the results at lowest values of the pion masses are
still large and currently do not allow to reach a definite conclusion. 
\vskip -0.1cm
\FloatBarrier

In order to extract the anomalous magnetic moment one needs to
 fit the $Q^2$-dependence of $G_M$. Typically one employs a dipole form 
to extrapolate at $Q^2=0$
introducing a model-dependence. Exploratory studies based  on a position
space method can yield $G_M(0)$ directly without having to perform a fit.
This method involves  taking the derivative of the relevant
correlator with respect to the momentum, allowing access to zero
momentum data. Thus, there is no need to assume a functional form for
the momentum dependence. Such a study is performed  for $G_M$~\cite{ETMC14c}
and the results are  shown in the right panel of Fig.~\ref{fig11}.  

%\newpage
\vspace{-0.4cm}
\subsection{Quark momentum fraction}
\vskip -0.2cm

\begin{figure}[!h]
\cl{\includegraphics[scale=0.285]{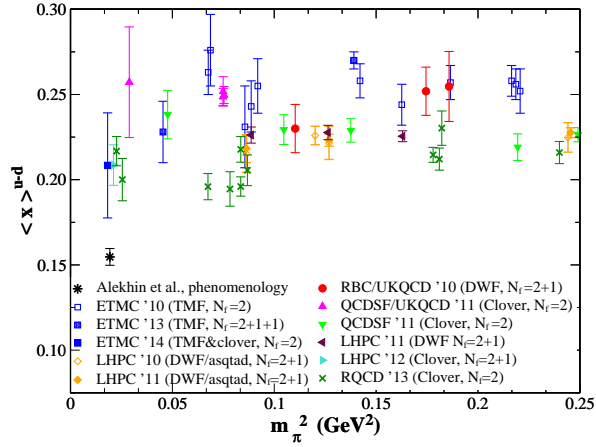}}
\vskip -0.3cm
\caption{Recent lattice results on the isovector unpolarized
  moment, $\langle x \rangle_{u-d}$, as a function of the squared
pion mass. The lattice data correspond to: 
$N_f{=}2$ TMF (ETMC~\cite{Alexandrou:2011nr}), 
$N_f{=}2{+}1$ DWF on asqtad sea (LHPC~\cite{Bratt:2010jn,Syritsyn:2011vk}),
$N_f{=}2{+}1$ DWF (RBC/UKQCD~\cite{Aoki:2010xg}, LHPC~\cite{Syritsyn:2011vk}),
$N_f{=}2$ Clover (QCDSF/UKQCD~\cite{Pleiter:2011gw}, QCDSF~\cite{Sternbeck:2012rw}, RQCD~\cite{Bali:2013nla,RQCD14}),
$N_f{=}1{+}2$ Clover (LHPC~\cite{Green:2012ud}), 
$N_f{=}2{+}1{+}1$ TMF (ETMC~\cite{Alexandrou:2013joa}),
$N_f{=}2$ TMF with Clover (ETMC~\cite{ETMC14}). 
The phenomenological value shown is 0.1646(27)
\cite{Alekhin:2009ni}.}
\label{fig13}
\end{figure}

\begin{figure}[!h]
\cl{\includegraphics[scale=0.19]{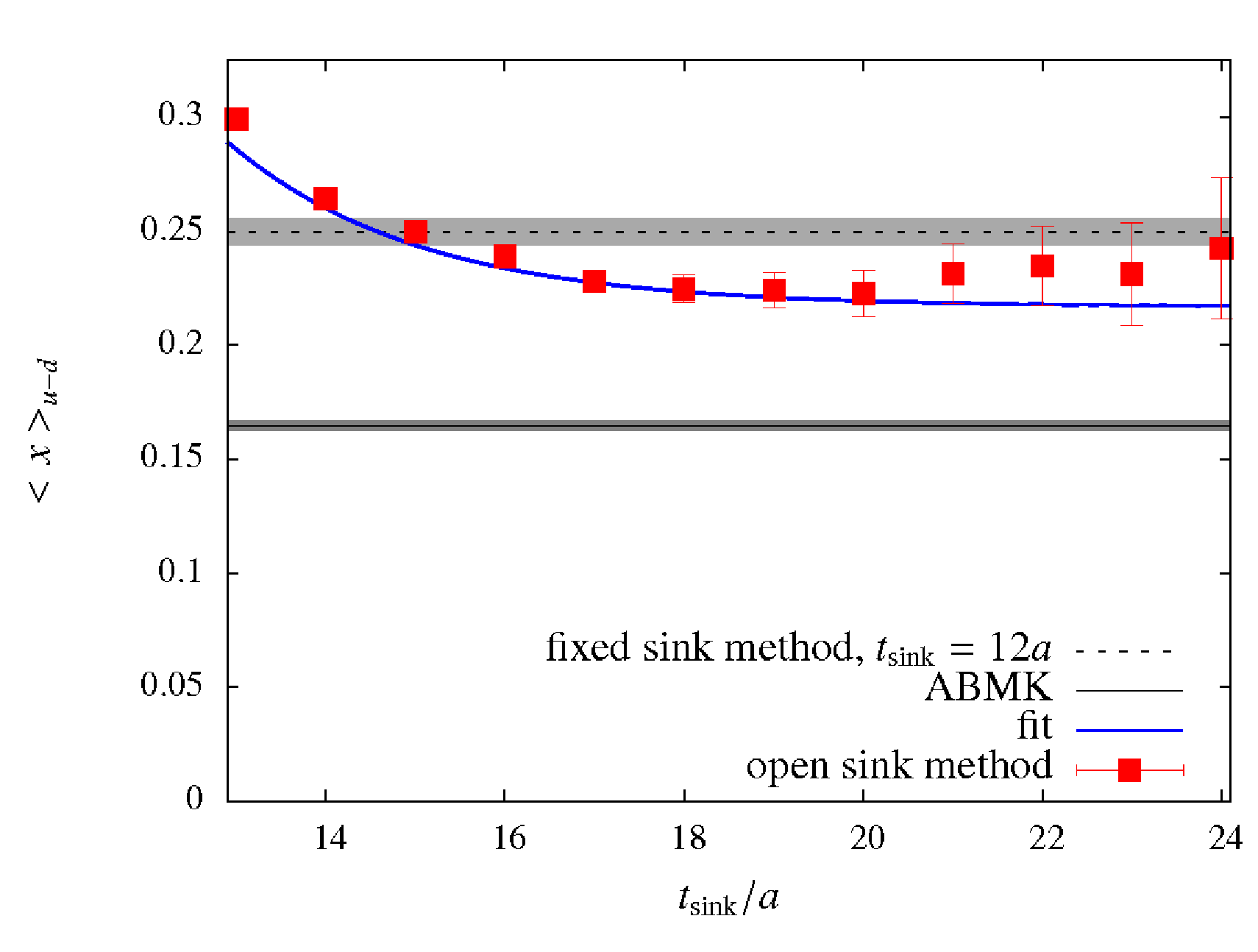}\,\,
 \includegraphics[scale=0.49]{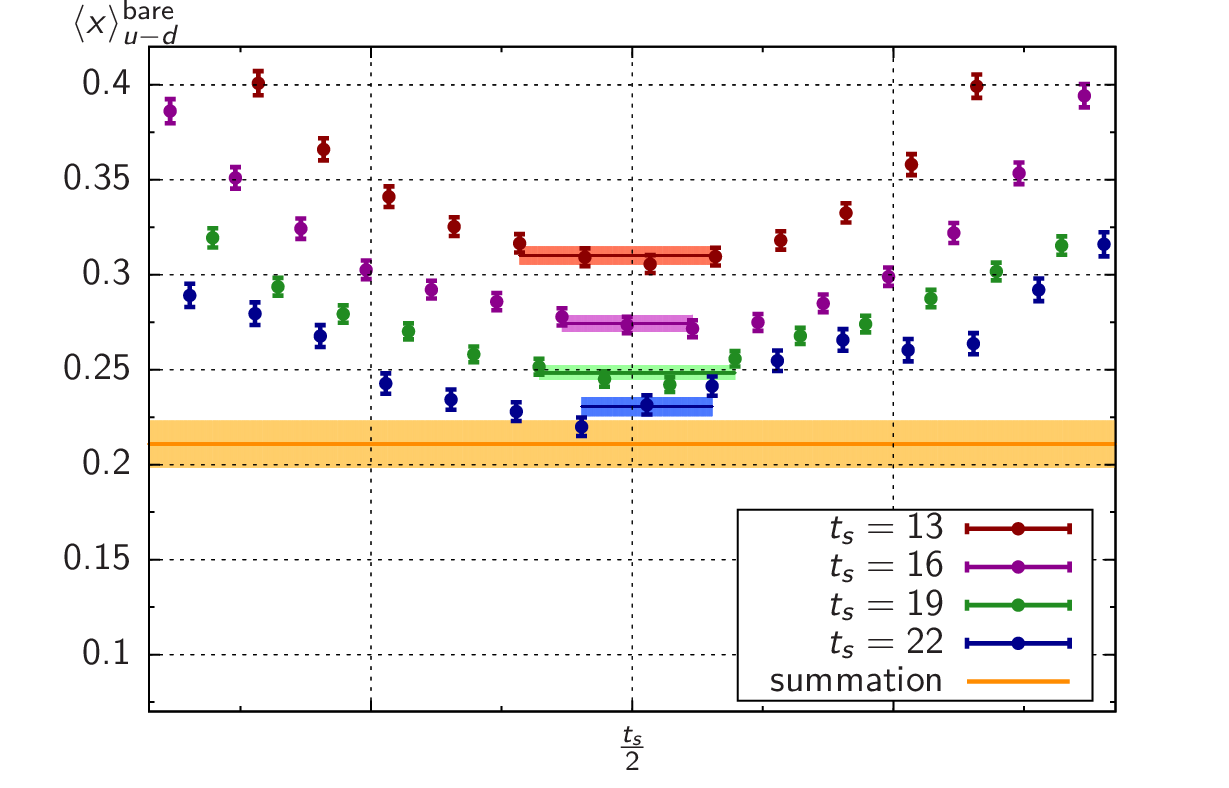}}
\vskip -0.1cm
\cl{\hspace{1cm} (a) \hspace{5.5cm} (b)\hspace{1cm}}
\vskip -0.05cm
\cl{\includegraphics[scale=0.49]{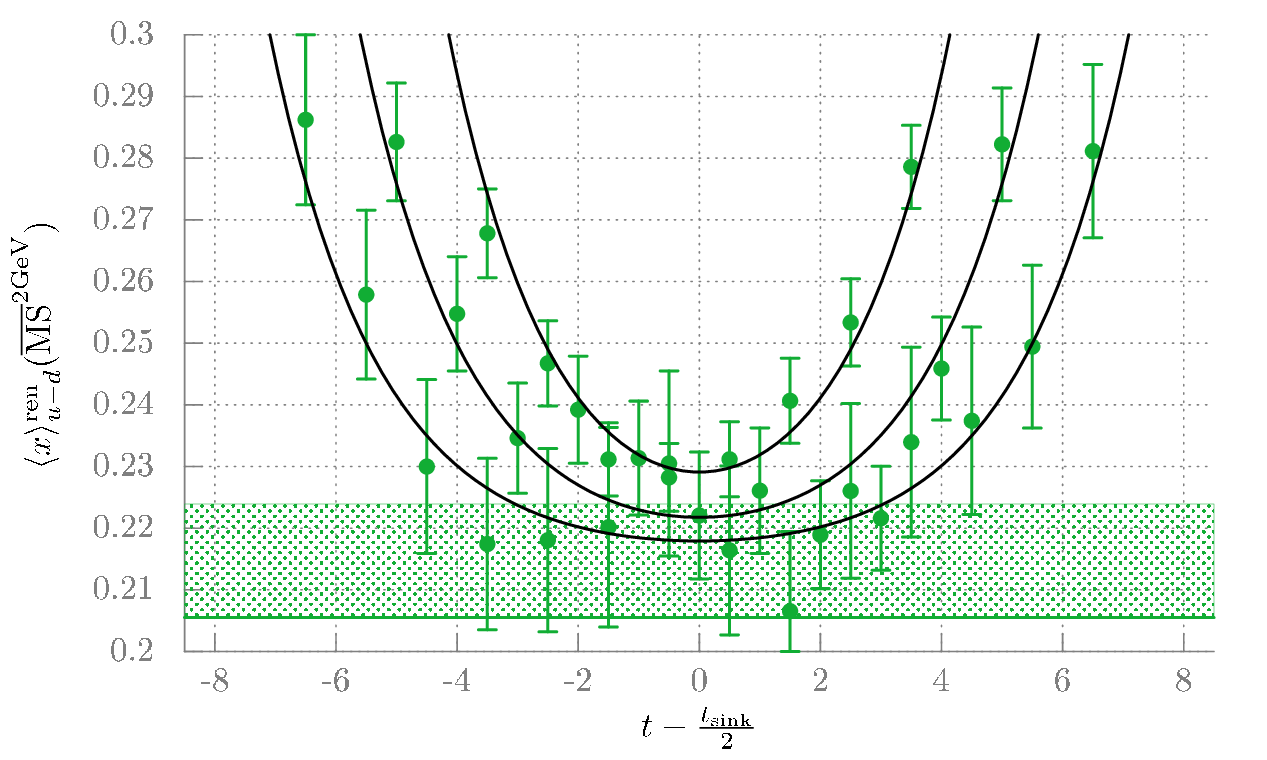}\,\,
\includegraphics[scale=0.19]{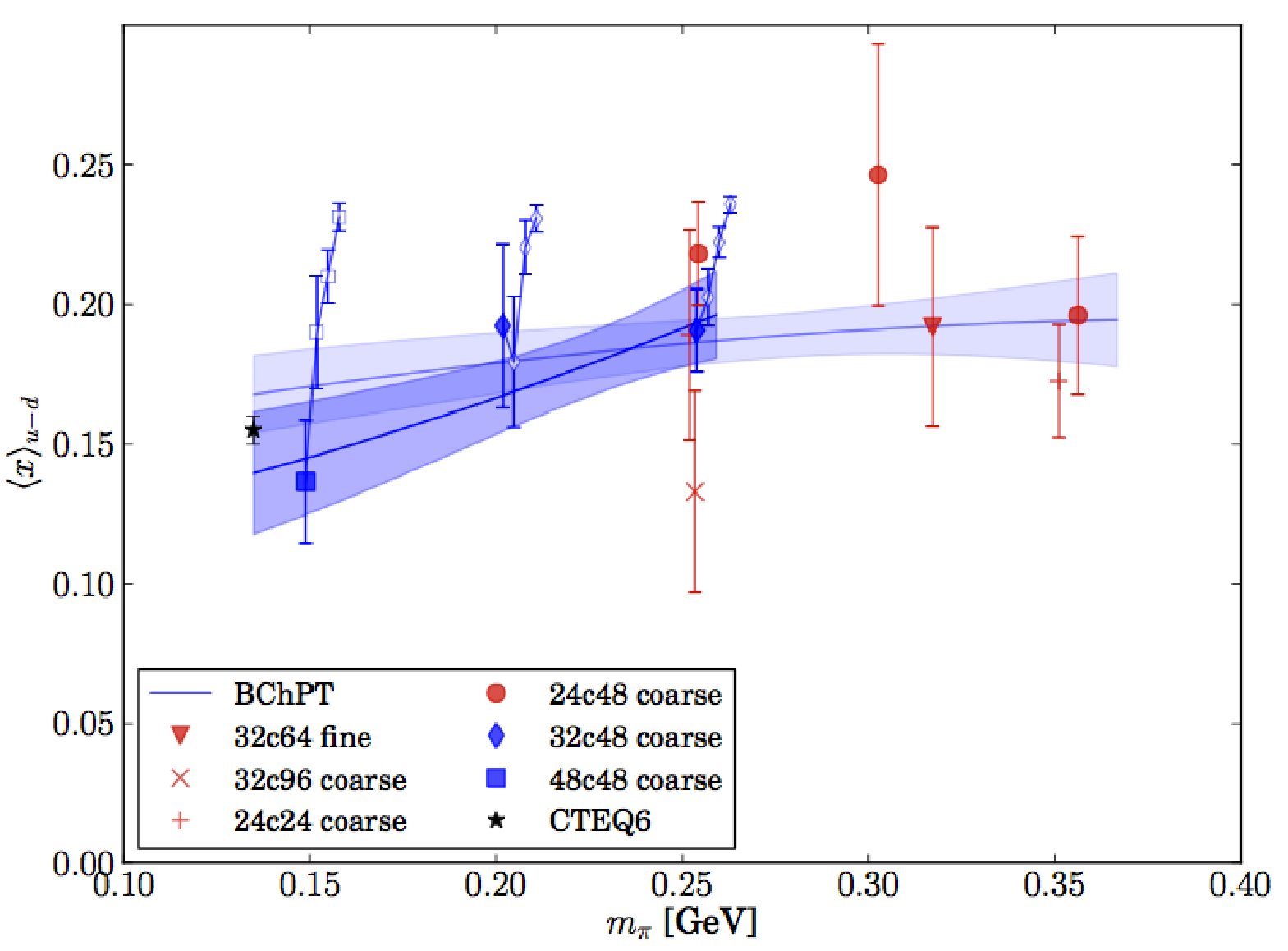}}
\vskip -0.1cm
\cl{\hspace{1cm} (c) \hspace{5.5cm} (d)\hspace{1cm}}
\vskip -0.2cm
\caption{Studies for excited states contaminations using: (a) TMF at
$m_\pi=373$ MeV (ETMC~\cite{Dinter:2011sg}), (b) Clover at
$m_\pi=340$ MeV (MAINZ~\cite{Jager:2013kha}), (c) Clover at
$m_\pi=150$ MeV (RQCD~\cite{RQCD14}), and (d) Clover at $m_\pi=149$
MeV (LHPC~\cite{Green:2012ud}).}
\label{fig14}
\end{figure}

Another important observable of hadron structure is the quark momentum
fraction. This is directly related to the first moment of the unpolarized structure
function through the operator product expansion. It can be extracted 
from  the forward matrix element of the one-derivative vector
current, and it is a scheme and scale dependent quantity.
% and it is universaly
%converted to $\overline{\rm MS}$ scheme at a scale of
%2GeV.
Experimentally it is measured in Deep Inelastic Scattering  where
phenomenological input is required in order to extract it from
measurements. Fig.~\ref{fig13} shows results on the
isovector momentum fraction converted to $\overline{\rm MS}$ scheme at
a scale of 2~GeV. It has been known for some time that the lattice QCD
value using different discretizations is larger than the experimental
one. It is worth mentioning that the phenomenological value
extracted from different analyses (see Refs. [56-61] of
Ref.~\cite{Alexandrou:2013joa}) shows a spread, which, however, is
significantly smaller than the discrepancy shown by the lattice
data. Nevertheless, lattice results close to the physical point
obtained from different discretizations are in agreement.
These correspond to source-sink separation of $\sim$ 1-1.2 fm which,
as discussed below, might not be large enough.
\vskip -0.14cm
\FloatBarrier
\vskip -0.13cm
\noindent
A number of studies were undertaken to examine the role of excited
states in the extraction of $\langle x \rangle_{u-d}$ and all works
corroborate that this observable suffers from excited states
contamination, as demonstrated in Fig.~\ref{fig14}. Various pion
masses have been explored (149-373 MeV), as well as the plateau and
summation (Figs.~\ref{fig14}b - \ref{fig14}d)
methods. The data of Figs.~\ref{fig14}a - \ref{fig14}d correspond
to lattice spacings 0.082, 0.05, 0.06, 0.116 fm, respectively and the
source-sink separation for each plot varies within the range:
[0.94-1.87], [0.6-1.4], [0.63-1.05] and [0.9-1.4] fermi.
It is shown that excited states contamination are accounted for a
decrease of about 10$\%$ in the value of $\langle x \rangle_{u-d}$ as
compared to the value extracted using sink-source separations of about
1 fm. Thus, one should seek for convergence with increasing the
source-sink separation and agreement with the summation method. Only
then we will have confidence in the final result.

\vskip 0.1cm
Another potential source of systematic error is the
renormalization functions which should be determined non-perturbatively. The
scheme which is particularly convenient is the RI$'$ defined as:
\vspace{-0.13cm}
\be
Z_q = \frac{1}{12} {\rm Tr} [\left(S^L(p)\right)^{-1}\, S^{\rm
    Born}(p)] \Bigr|_{p^2=\bar\mu^2},\quad 
Z_q^{-1}\,Z_{\cal O}\,\frac{1}{12} {\rm Tr} [\Gamma^L_{\cal
    O}(p) \,\left(\Gamma^{\rm Born}_{\cal O}(p)\right)^{-1}] \Bigr|_{p^2=\bar\mu^2}=1\,.
\ee

\newpage
\FloatBarrier

\begin{figure}[!h]
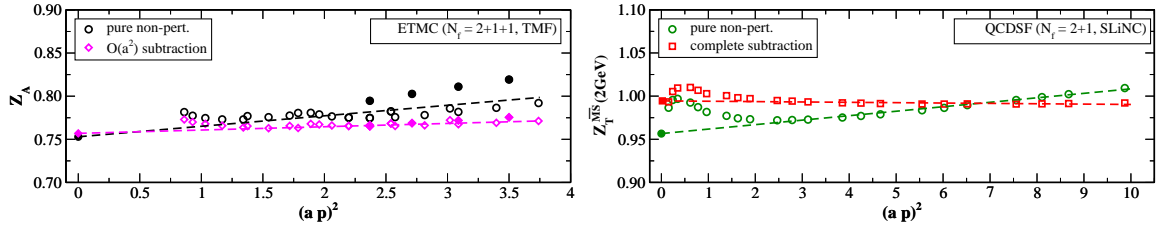

\cl{\includegraphics[scale=0.28]{./Za_ETMC_Oa2.eps}\,
    \includegraphics[scale=0.28]{./Zt_QCDSF_allOa.eps}}
\vskip -0.3cm
\caption{Left (Right): $Z_A$~\cite{Alexandrou:2012mt}
  ($Z_T$~\cite{Constantinou:2014fka}) as a function of the renormalization scale.}
\label{fig12}
\end{figure}

 For the scheme and scale
dependent operators such as  $\langle x \rangle$, we need to
convert to the $\overline{\rm MS}$ at 2~GeV, a step that also 
potentially can introduce systematic errors. We find that any error
is insignificant when one utilizes
the NNLO formulas.  Lattice cut-off effects can also be ameliorated by using
two alternative approaches: the
first one employed by ETMC~\cite{Alexandrou:2012mt} is based on the
subtraction of ${\cal O}(a^2)$ terms computed perturbatively
\cite{Constantinou:2009tr}, which improves the quality of the plateau
(see left panel of Fig~\ref{fig12}). The second approach employed by
QCDSF~\cite{Constantinou:2014fka} is the subtraction of the one-loop
artifacts to all order in the lattice spacing, eliminating the slope
of the plateau with respect to $(a\,p)^2$. It was shown that the
complete subtraction works very well even at high scales where one might have expected the
 ${\cal O}(a^2)$ subtraction  not to be sufficient. 
\vskip -0.15cm
\FloatBarrier
\vskip -0.15cm
One may also control {\it a priori} the lattice artifacts by using momenta
that have small non-Lorentz invariant contributions: empirically
{\small{$(\sum_\rho p_\rho^4)^4/(\sum_\rho p_\rho^2)^2 \langle
    \,0.4$}}. For example, the black filled circles shown in
Fig.~\ref{fig12} for the case of $Z_A$ violate the above condition and
are consistently higher than the rest. Note that, upon subtraction
of the ${\cal O}(a^2)$ terms, the magenta points fall nicely on one line.
Thus it is apparent that a
combination of perturbative and non-perturbative methods
can improve the determination of the renormalization factors.

%\newpage
\vskip -0.4cm
\subsection{Nucleon Spin}

\vskip -0.2cm
A long-standing puzzle has been the spin fraction carried by the quarks in the
nucleon, which is found to be about half of the total nucleon
spin~\cite{Ashman}. It was proposed that gluons in a polarized proton
would carry a fraction of the spin, which however would be unnaturally
large if it were to resolve the spin crisis. It is now understood that
the resolution of this puzzle requires to take into account the
non-perturbative structure of the proton. Using the lattice QCD
formalism one can provide significant input towards understanding this
open issue. The total nucleon  spin  is generated by the quark
orbital angular momentum ($L^q$), the quark spin ($\Sigma^q$) and the
gluon angular momentum ($J^G$). The quark components are related to
 $g^q_A$ and the GFFs of the one-derivative vector at $Q^2=0$:
\vspace{-0.24cm}
\be
\frac{1}{2} = \sum_q \left(L^q + \frac{1}{2}\Delta\Sigma^q \right) + J^G\,,
\quad J^q = \frac{1}{2} \left( A_{20}^q + B_{20}^q  \right)\,,  \quad L^q=J^q-\Sigma^q \,,  \quad \Sigma^q = g_A^q\, .
\ee 
\vskip -0.24cm
\noindent
Since we are interested in the individual quark contributions to the
various components of the spin, one needs to consider the
disconnected contributions. 

The computation of disconnected diagrams
using improved actions with dynamical fermions became feasible over
the last years and for the proper renormalization of the individual quark
and isoscalar contributions one should take into account the singlet
operator. Non-perturbatively these are unknown (for a recent study
using the Feynman-Hellmann approach, see Ref.~\cite{Chambers:2014pea}), 
but are expected to be small since in perturbation theory they first
appear to two loops. Recent results by the Cyprus
group~\cite{Cyprus14} show that only the scalar and axial
contributions survive to two loops. The results are general enough to
be used by any combination between Wilson, Clover, SLiNC, TM fermions
and any of the Symanzik improved gluons. 
The difference $Z_{\cal O}^{singlet} - Z_{\cal O}^{non-singlet}$ may
be added to the non-perturbative non-singlet estimates for the
Z-factors before using them to renormalize the disconnected
contributions.
\vskip -0.2cm
\begin{figure}[!h]
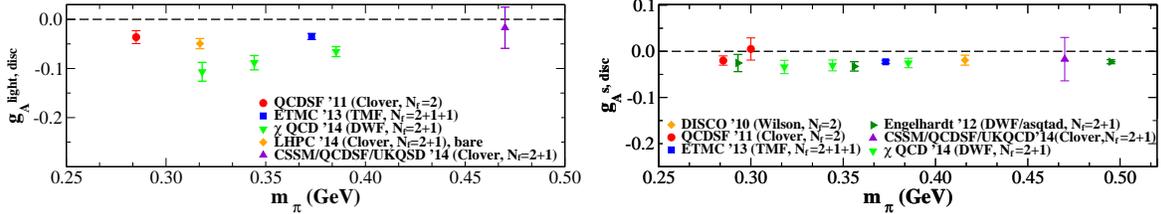

\cl{\includegraphics[scale=0.27]{./gA_disc_light_vs_mpi.eps}\,\,
\includegraphics[scale=0.27]{./gA_disc_strange_vs_mpi.eps}}
\vskip -0.4cm
\caption{Disconnected contributions for $g_A^q$ for the light (left) and
  strange (right) quark contributions.}
\label{fig15}
\end{figure}
\FloatBarrier
\vskip -0.2cm
A number of results have appeared recently
where the disconnected loop contributions
to
$g_A$ are evaluated as shown in Fig.~\ref{fig15}. We
observe a nice agreement among results using a number of methods  both
for the light~\cite{QCDSF:2011aa,Abdel-Rehim:2013wlz,chiQCD14b,LHPC14,QCDSF14} as well as for 
the strange quark contributions~\cite{Babich:2010at,QCDSF:2011aa,Abdel-Rehim:2013wlz, 
Engelhardt:2012gd, LHPC14,QCDSF14}. For $g_A^{light}$ we find
$\sim10\%$ contributions compared to the connected part that must be
taken into account in the discussion of the spin carried by quarks in
the proton. These contributions are negative and thus reduce the value of $g_A^q$.
 There is also a computation for the disconnected contribution to $\langle x
\rangle$ using TM fermions~\cite{Abdel-Rehim:2013wlz} at $m_\pi=373$
MeV, which is also of interest since it contributes to the spin. At
this particular pion mass it was found to be compatible with zero.

\begin{figure}[!h]
\cl{\includegraphics[scale=0.2,angle=90]{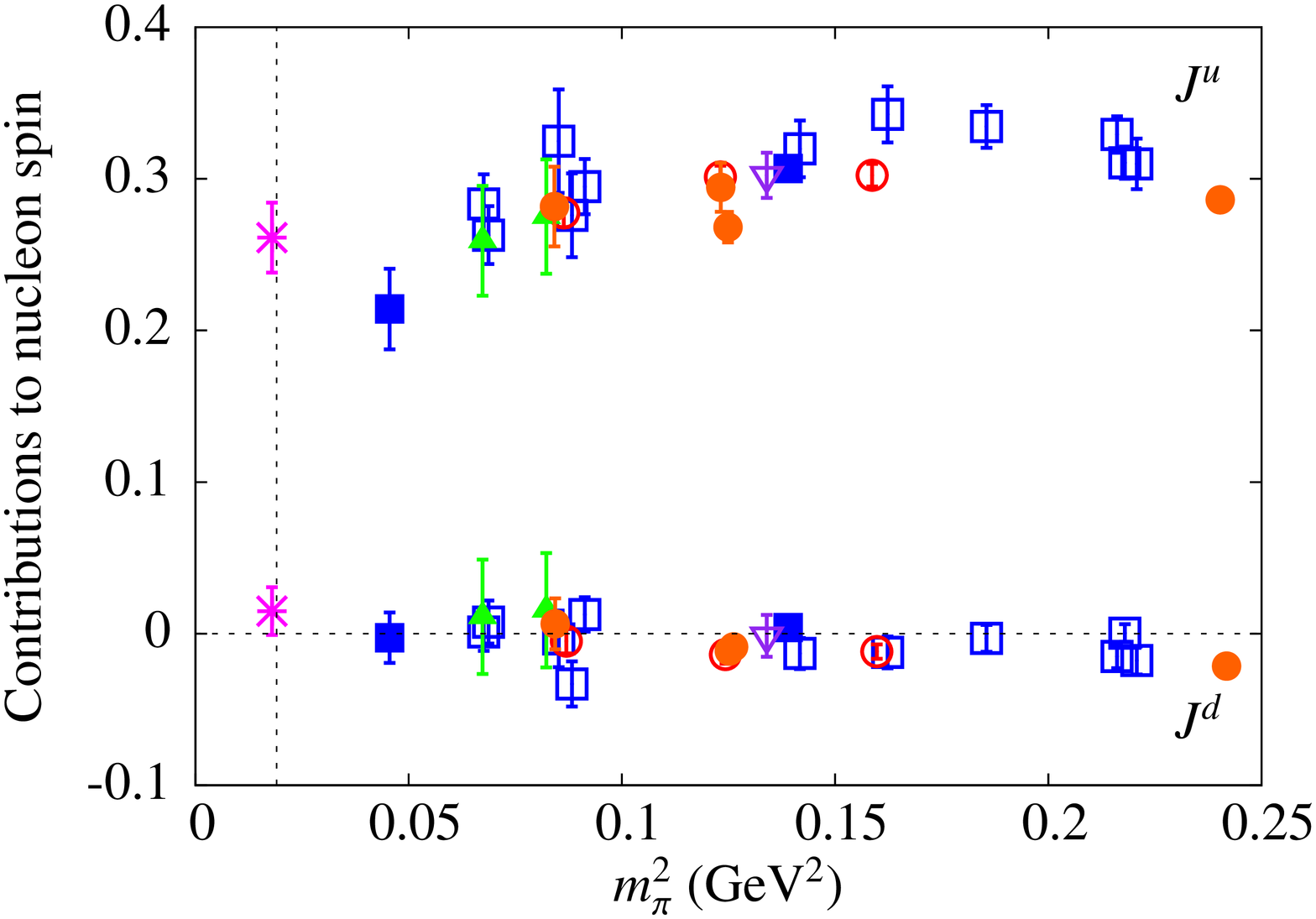}\,\,
    \includegraphics[scale=0.2,angle=90]{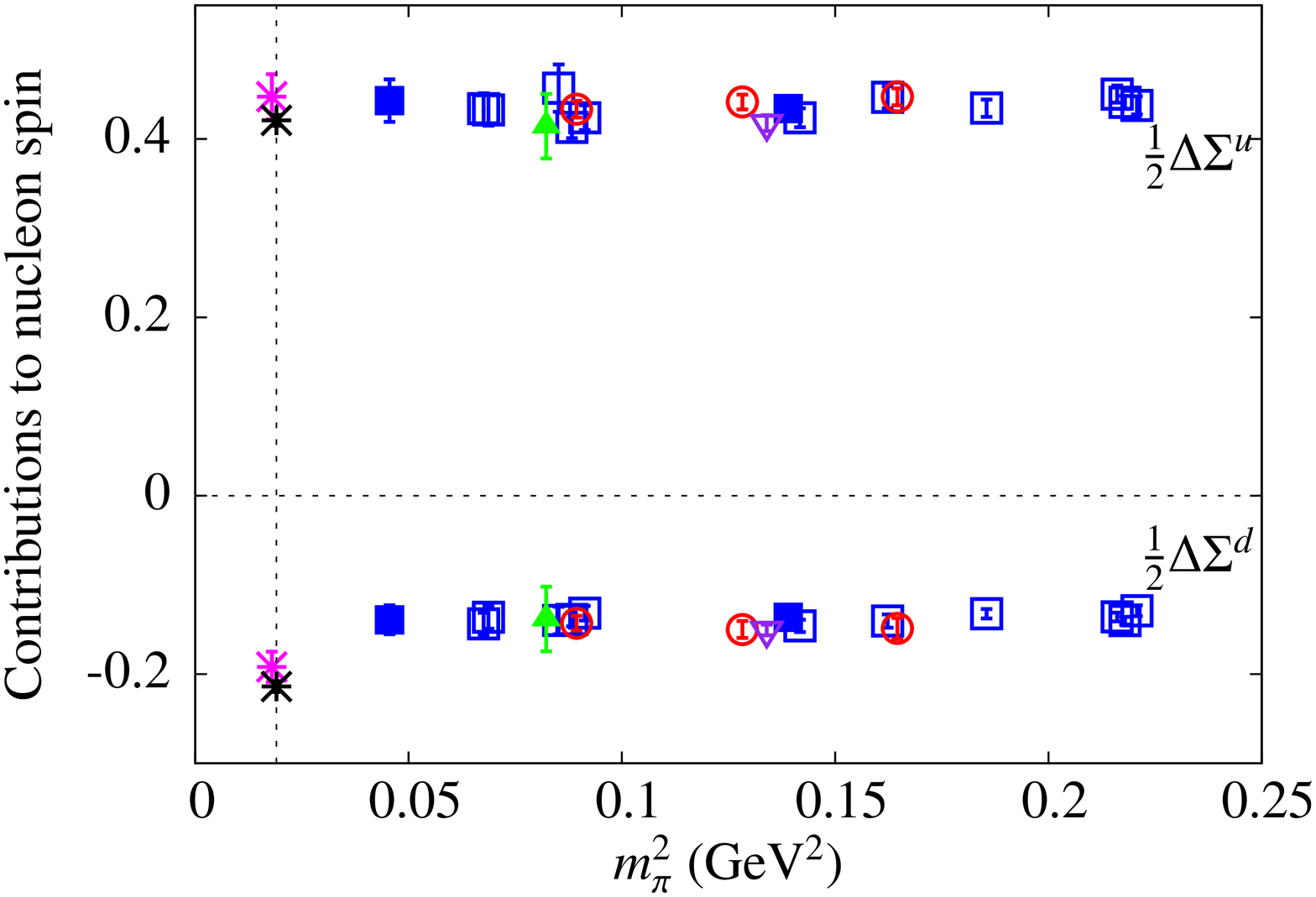}}
\vskip -0.45cm
\caption{The total spin, $J^q$, and the quark spin, $\Sigma^q$,
  carried by the up and down quarks.The lattice data correspond to:
$N_f{=}2{+}1$ DWF and DWF on asqtad (LHPC~\cite{Syritsyn:2011vk}),
$N_f{=}2$ Clover (QCDSF/UKQCD~\cite{Sternbeck:2012rw}),
$N_f{=}2$ TMF (ETMC~\cite{Alexandrou:2011nr})
$N_f{=}2{+}1{+}1$ TMF (ETMC~\cite{Alexandrou:2013joa})
$N_f{=}2$ TMF with Clover (ETMC~\cite{ETMC14}). }
\label{fig16}
\end{figure}

In Fig.~\ref{fig16} we show results on the total spin, $J^q$.
It is apparent that the  u-quark  exclusively carries
the spin attributed to the quarks in the nucleon since $J^d$ is
consistent with zero for all pion masses and lattice
discretization schemes.
 The quark distribution to the intrinsic spin in also
shown in Fig.~\ref{fig16}. There is a nice agreement
between the  results at the physical pion mass using TM
fermions~\cite{ETMC14} and the experimental values for both the u- and
the d-quarks. In all the TMF data, the $Z_A^{singlet}$ is computed
perturbatively~\cite{Cyprus14} and has been included in the
determination of the results. The disconnected contributions
have been neglected from most data except for one TMF ensemble at
$m_\pi=373$ MeV. The effect is shown by the deviation of the filled
blue square which has no disconnected contributions from the
violet triangle, which include them. Although the effect is small it
is larger than the statistical error and thus one needs to take them
into acccount. The lattice results thus corroborate the missing spin
contribution arising from the quarks. Whether gluonic degrees of
freedom are responsible for the other half is debatable and remains an
open question.
\vskip -0.35cm
\FloatBarrier

\vskip -0.3cm
\section{Hyperon Form Factor}
\vskip -0.2cm
\subsection{Electromagnetic Form Factors}

\begin{figure}[!h]
\cl{\includegraphics[scale=0.19]{./GE_GM_Sigma_IV_CSSM_QCDSF_UKQCD_2.eps}\quad
\includegraphics[scale=0.48]{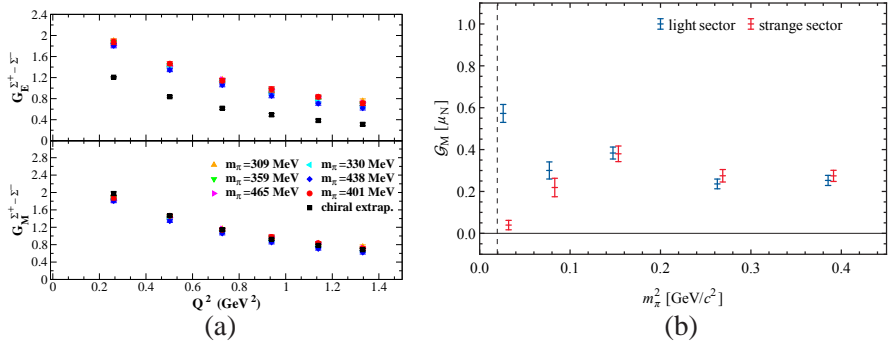}}
\vskip -0.2cm
\cl{\hspace{1cm} (a) \hspace{5.5cm} (b)\hspace{1cm}}
\vskip -0.22cm
\caption{(a). $G_E^\Sigma$ and $G_M^\Sigma$ as a function of $Q^2$ for
$m_\pi$ in the range of 309~MeV to 465 MeV along with the chiral extrapolation
\cite{Shanahan:2014uka,Shanahan:2014cga}. (b). The light and strange
quark contributions to $G_M$ of the $\Lambda(1405)$~\cite{Menadue:2011pd,CSSM14}.} 
\label{fig17}
\end{figure}

The form factors of baryons other than the nucleon are poorly known
and not amenable to measurements. Thus lattice QCD  can provide
significant input on these quantities. In a recent work of 
CSSM/QCDSF/UKQCD~\cite{Shanahan:2014uka, Shanahan:2014cga} the hyperon
EM FFs were computed at various values of $m_\pi$. 
By setting the electric charge of the sea quarks to zero,
disconnected quark loops are completely removed. The authors perform
independent chiral fits to the data at each value of $Q^2$, and thus,
the coefficients in the chiral expansion are considered to be the
chiral limit at fixed $Q^2$. This procedure is shown for the $G_E$ and
$G_M$ of the $\Sigma$ in Fig.~\ref{fig17}a, where the chirally
extrapolated results are shown with black points. For the case of the
proton, the authors find agreement with the experimental values after
performing a chiral extrapolation using their approach. 
\vskip -0.25cm
\FloatBarrier
CSSM has studied the magnetic FF of the lowest lying excitation
of the $\Lambda$ baryon, $\Lambda(1405)$, which despite its quark
structure that includes a strange quark is lighter than other excited
spin 1/2 baryons. There are speculations that $\Lambda(1405)$ is a
superposition of molecular meson-baryon states where the $\pi\,\Sigma$
and $\bar K\,N$ channels play a significant role. Study of the strange
contribution to $G_M$ may shed light on the internal structure of this
particle, since if it is an $\bar{K}\,N$ bound state, then the s-quark
is confined in a spin zero kaon with no preference in the spin
orientation, and thus, $G_M \to 0$. The lattice data for this study
were extracted using a variational analysis for several values of the
pion mass. Simulations very close to the physical point show a
significant drop of $G_M^s$ which indeed becomes almost zero (see
Fig.~\ref{fig17}b). According to the authors, this is an indication
that $\Lambda(1405)$ is dominated by a $\bar K\,N$ bound state.

\subsection{Axial Charges of Hyperons}

The axial charge of hyperons is another example where methods used in
the nucleon sector can give predictions needed, for instance, as an
input in chiral effective theories.   On the lattice, the
hyperon axial charges are extracted from the matrix
element
$\langle B(p') | \bar\psi(x)\,\gamma_\mu\,\gamma_5\,\psi(x)|B(p)\rangle\,\Big{|}_{q^2=0}$.

\begin{figure}[!h]
\cl{\includegraphics[scale=0.56]{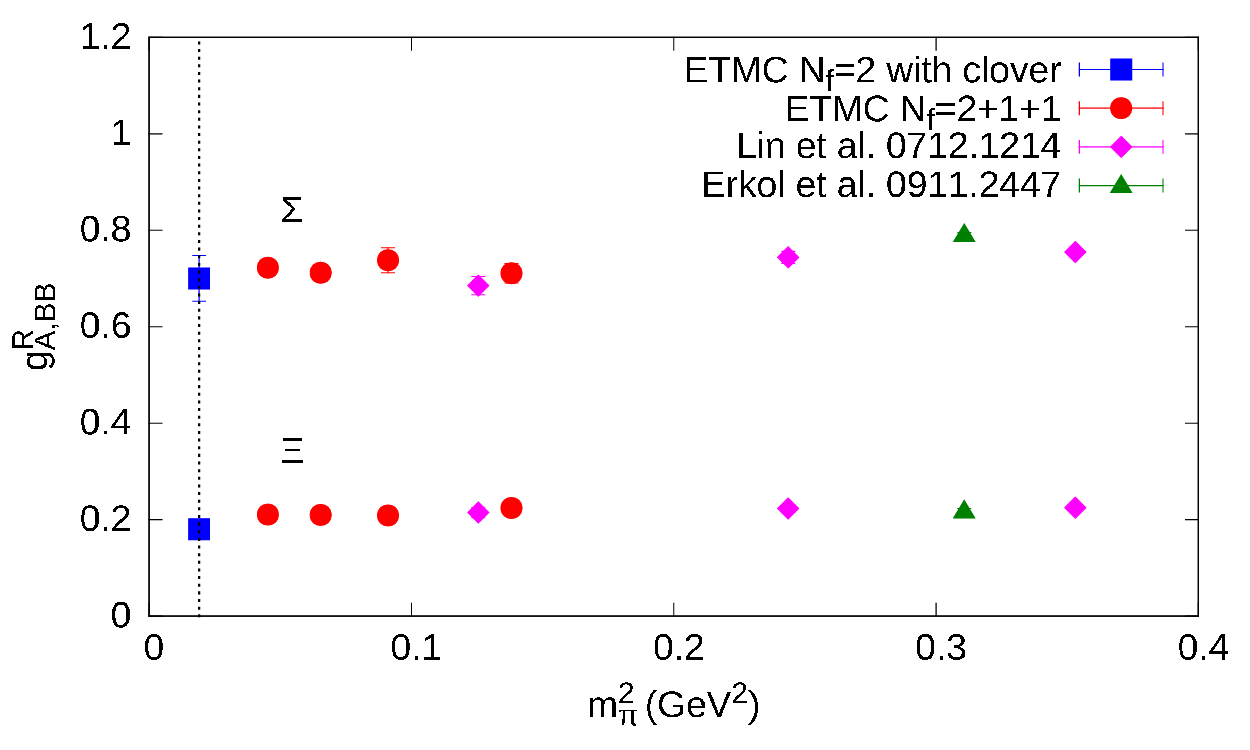}\,\,
    \includegraphics[scale=0.54]{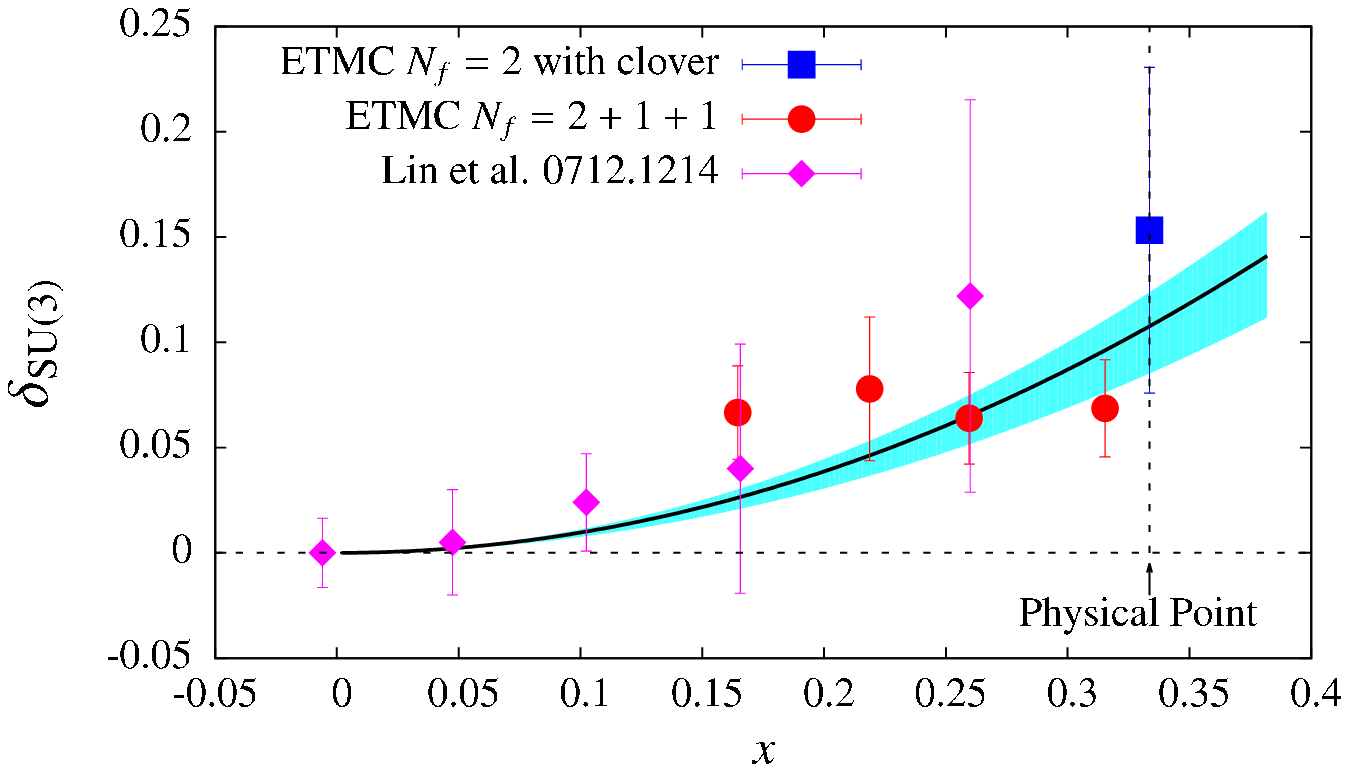}}
\vskip -0.2cm
\caption{Left: $g_A$ for $\Sigma$ and $\Xi$ as a function of $m_\pi^2$
(\cite{ETMC14} and references therein). Right: SU(3) breaking parameter $\delta_{SU(3)}$ versus
$x$.}
\label{fig18}
\end{figure}

In  Fig.~\ref{fig18} we show $g_A$
for $\Sigma$ and $\Xi$ for different discretizations as a function of $m_\pi$.
Results using TMF probe the small pion mass region reaching the physical point.
There is no strong pion mass dependence even for pion masses close to the
phsyical value. This behaviour is also observed in the case of other baryons.
In the same figure we show the  SU(3) symmetry breaking $\delta_{SU(3)}=g_A^N-g_A^\Sigma+g_A^\Xi$  versus $x=\left(m_K^2-m_\pi^2\right)/(4\pi^2f_\pi^2)$. As expected it increases quadratically with $x$. 
Simulations at the physical point with TMF~\cite{ETMC14} show about 15$\%$
SU(3) flavor  breaking.
\vskip -0.2cm
\FloatBarrier

\vspace{-0.45cm}
\section{Mesons}
\vskip -0.1cm
Apart from studies of baryon structure, there are recent
developments in the meson sector, which however we will only
summarize here, focusing on the pion and the 
$\rho$-meson.  
\vskip -0.1cm
\subsection{Pion Quark Distribution Function}

The pion is a Goldstone boson playing a central role in
 chiral symmetry breaking with many consequences in nuclear and particle physics phenomena. Although it is
consider a simple bound state its internal structure in not well
studied. In fact, information about pion parton distribution functions
(PDFs) rely on input from nucleon PDFs. Thus, Lattice QCD is in a
unique position to provide results from first principles. 

\begin{figure}[!h]
\cl{\includegraphics[scale=0.26]{./aver_x_q_pion_vs_mpi2_v2.eps}\quad
\includegraphics[scale=0.405]{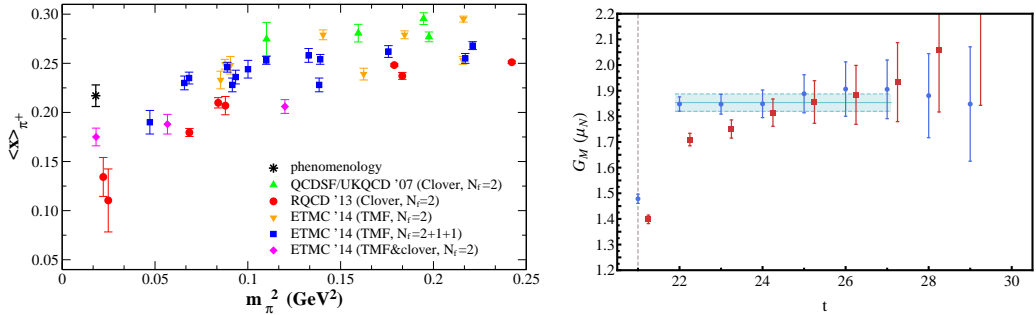}}
\vskip -0.3cm
\caption{Left: $\langle x \rangle_{\pi^+}$ as a function of $m_\pi^2$ for: 
$N_f{=}2$ Clover (QCDSF/UKQCD~\cite{Brommel}, RQCD~\cite{Bali:2013gya}),
$N_f{=}2$ TMF (ETMC~\cite{Baron:2007ti}),
$N_f{=}2{+}1{+}1$ TMF (ETMC~\cite{ETMC14d}),
$N_f{=}2$ TMF with Clover (ETMC~\cite{ETMC14d}). 
Right: Plateau for $G_M$ of the $\rho$-meson~\cite{CSSM14b} for the
standard (red points) and variational (blue points) methods.}
\label{fig19}
\end{figure}

In a recent study with TMF the momentum fraction  $\langle x \rangle$, in the pion is computed using the operator ${\cal O}_{44}$
\vspace{-0.27cm}
\be
{\cal O}_{44}(x) = \frac{1}{2}\bar u(x)[\gamma_4\,\Dlr_4 -\frac{1}{3}\sum_{k=1}^3\gamma_k\,\Dlr_k]u(x),\quad
\langle x \rangle_{\pi^+}^{\rm bare} = \frac{1}{2\,m_\pi^2}\,\langle\pi,\vec{0}|{\cal O}_{44}|\pi,\vec{0}\rangle .
\ee
\vskip -0.13cm
\noindent
The matrix element of this operator  is non-zero at
$Q^2=0$ and thus $\langle x \rangle$ can be easily extracted without requiring 
extrapolating the results to zero momentum transfer.
A stochastic time  source is utilized allowing  high
statistical accuracy at small computational cost.
 In the left panel of Fig.~\ref{fig19} the momentum fraction is plotted against
$m_\pi^2$ for Wilson TMF and Clover fermions. There is quite a spread
in the results obtained using the different discretizations that
may point to large lattice artifacts that need to be investigated.
Lattice results close to the physical point underestimate the phenomenological
value of $\langle x \rangle_{\pi^+}=0.0217(11)$~\cite{Wijesooriya:2005ir}. 
A comprehensive study of lattice artifacts is called for in order to 
understand the observed discrepancies in the lattice data.
\vskip -0.2cm
\FloatBarrier

\vskip -0.4cm
\subsection{$\rho$-meson Electromagnetic Form Factor}
\vskip -0.1cm

The EM of the $\rho$-meson are computed by the CSSM collaboration
using $N_f{=}2{+}1$ Clover fermions~\cite{CSSM14b}.
Since the $\rho$ is a resonance its form factors are not known.
The $\rho$-meson matrix element of the EM current can be
decomposed into a charge, magnetic and quadrupole form factor,
denoted by $G_C,\,G_M$ and $G_Q$, respectively. A variational approach
has been utilized and was found to separate excited states
effectively. The technique is applied on a set of
operators built from various source and sink smearings, applied to the
$\rho$ interpolating field $\chi^i_\rho(x)=\bar d(x) \gamma^i\,u(x)$. 
Four levels of Gaussian smearing were employed, thus,
a $4\times 4$ correlation matrix was analyzed. The authors find
substantial improvement in the determination of  $G_M$ and $G_Q$ using
the variational approach. In the right plot of Fig.~\ref{fig19} we
show the results for $G_M$ using the standard method and the
variational method. As can be seen, the plateau using the variational
method begin only one time slice afte the current insertion, as
compared to the standard method where no clear plateaus were observed.

\vskip -0.2cm
\section{Conclusions and Perspectives}
\vskip -0.12cm

Several major improvements in  algorithm  and techniques
coupled with increase in the computational power have allowed  lattice
QCD simulations with light quark masses fixed to their physical value.
Although this is a big achievement, many challenges lie ahead: 
development of appropriate algorithms to reduce the statistical errors
at reduced cost,  understanding how to treat unstable particles and
resonances, inclusion of  multi-particle states and computing
accurately observables that probe beyond the standard model physics
are some of them.

For hadron structure, simulations at different lattice spacings and
larger volumes are crucial for a proper study of lattice artifacts in
order to provide reliable results at the continuum limit. Such studies 
require an accuracy which is difficult to achieve with standard methods.
All-Mode-Averaging or other noise reduction techniques are thus
essential in order to settle some of the long-standing discrepancies 
reviewed in this talk. 

Similarly techniques developed for the computation of disconnected quark
loop diagrams, such as the truncated solver method~\cite{Bali:2009hu}
need to be improved since they become inefficient at the physical
point~\cite{ETMC14e}. Hierarchical probing proved to be very promising in
the evaluation of the electromagnetic form factors at pion mass of $\sim$300~MeV 
is doubtful that it can work so well at the physical point.
Thus, new ideas will be needed to compute disconnected contributions to 
hadron structure to an accuracy of a few percent.
This is a real challenge if lattice QCD wants to contribute in the debate 
on the charge radius of the proton.
 Utilization of new computer architectures such as  GPUs has proved essential
for the evaluation of disconnected diagrams and this is a  direction
that should be  pursued in the future.  

Other open issues such as  the nucleon spin may need the computation
of challenging quantities such as the gluonic contributions, which 
beyond gauge noise, are difficult to renormalized mixing with other operators.
Extending the formalism from the nucleon to  other baryons will be
another challenge since most of these particles decay strongly.

Despite the aforementioned challenges lattice QCD has entered a new era
in terms of  simulations. Having gauge configurations at the
physical point has eliminated one of  the systematic error that was
inherent in all lattice calculations in the past, namely the difficult
to quantify systematic error due to the chiral extrapolation in
particular for the baryon-sector. This has opened the way to address a
number of challenging systems directly at the physical point. This
hold the promise to resolve discrepancies on benchmark quantities like
$g_A$  and reliably compute quantities relevant for revealing possible
new physics.

\bigskip
{\bf {Acknowledgments:}} I would like to thank all those who have
shared with me their recent work (listed in alphabetic order):
C. Alexandrou, G. Bali, M. Gong, R. Gupta, C. Kallidonis, G. Koutsou,
D. Leinweber, K.-F. Liu, S. Meinel, S. Ohta, B. Owen, H. Panagopoulos,
Th. Rae, Ph. Shanahan, C. Urbach, Y.-B. Yang, and J. Zanotti. 
I greatly appreciate the fruitful discussions I had with my
collaborator C. Alexandrou while preparing this talk. Last but not
least I thank A. Abdel-Rehim, K. Hadjiyiannakou, G. Koutsou for
discussions on technical oriented topics.
The speaker is supported by the Cyprus Research Promotion Foundation
grand TECHNOLOGY/$\Theta$E$\Pi$I$\Sigma$/0311(BE)/16.

\end{document}